\documentclass[jcp,reprint,amsmath,amssymb,graphicx,longbibliography,superscriptaddress]{revtex4-1}

\draft 

\usepackage{graphicx}
\usepackage{dcolumn}
\usepackage{bm}

\usepackage[utf8]{inputenc}
\usepackage[T1]{fontenc}
\usepackage{mathptmx}
\usepackage{dsfont}
\usepackage{xcolor}
\usepackage{todonotes}
\usepackage[unicode]{hyperref}
\hypersetup{
   unicode=true,          
   plainpages=false,
   colorlinks=true,       
   citecolor=blue,        
}

\usepackage{url}

\begin{document}


\title{
Improved walker population control for full configuration interaction quantum Monte Carlo}




\author{Mingrui Yang}
\email{M.Yang4@massey.ac.nz}
\affiliation{%
New Zealand Institute for Advanced Study and Centre for Theoretical Chemistry and Physics, Massey University, Auckland 0632, New Zealand%
}%
\affiliation{%
Dodd-Walls Centre for Photonic and Quantum Technologies, Dunedin 9056, New Zealand%
}%
\affiliation{%
MacDiarmid Institute for Advanced Materials and Nanotechnology, Wellington 6140, New Zealand%
}%

\author{Elke Pahl}%
\email{Elke.Pahl@auckland.ac.nz}
\affiliation{%
Department of Physics, University of Auckland, Auckland 1010, New Zealand%
}%
\affiliation{%
MacDiarmid Institute for Advanced Materials and Nanotechnology, Wellington 6140, New Zealand%
}%
\affiliation{%
School of Natural and Computational Sciences, Massey University and Centre for Theoretical Chemistry and Physics, Auckland 0632, New Zealand%
}%
\affiliation{
 Max Planck Institute for Solid State Research, Heisenbergstra{\ss}e 1,
70569 Stuttgart, Germany
}%

\author{Joachim Brand}
\email{J.Brand@massey.ac.nz}
\affiliation{%
New Zealand Institute for Advanced Study and Centre for Theoretical Chemistry and Physics, Massey University, Auckland 0632, New Zealand%
}%
\affiliation{%
Dodd-Walls Centre for Photonic and Quantum Technologies, Dunedin 9056, New Zealand%
}%
\affiliation{
 Max Planck Institute for Solid State Research, Heisenbergstra{\ss}e 1,
70569 Stuttgart, Germany
}%


\date{\today}

\begin{abstract}
Full configuration interaction quantum Monte Carlo (FCIQMC)  is a stochastic approach for finding the ground state of a quantum many-body Hamiltonian. It is based on the dynamical evolution of a walker population in Hilbert space, which samples the ground state configuration vector over many iterations. Here we
present a modification of
the original protocol for walker population control of Booth \textit{et al.} JCP \textbf{131}, 054106 (2009)
in order to achieve equilibration at a pre-defined average walker number and to avoid walker number overshoots.
The dynamics of the walker population is described by a noisy damped harmonic oscillator and controlled by two parameters responsible for damping and forcing, respectively,
for which reasonable values are suggested. 
We further introduce a population growth witness that can be used to detect annihilation plateaus related to overcoming the FCIQMC sign problem.
Features of the new population control procedure such as precise walker number control and fast equilibration
are demonstrated.
The standard error of the shift estimator for the ground state energy
as well as the population control bias are
found to be unaffected by the population control procedure or its parameters.
The improved control of the walker number, and thereby memory consumption, is a desirable feature required for automating FCIQMC calculations and requires minimal modifications to existing code.
\end{abstract}


\maketitle 



%
%

%


\section{Introduction}

Quantum Monte Carlo methods have proven invaluable tools for providing accurate results for strongly-correlated quantum many-body problems in different areas of physics and chemistry \cite{Carlson2012,Ceperley1996,Foulkes2001,Needs2010}.
One of the most straightforward approaches to solve a quantum problem with a definite particle number is to build a matrix representation of the Hamiltonian in  a Fock basis, i.e.~the properly symmetrized or anti-symmetrized product wave functions
of  $N$ bosonic or fermionic quantum particles, respectively, in a finite number $M$ of single-particle modes. Diagonalizing this matrix to obtain the ground or excited quantum states is known as exact diagonalization or full configuration interaction \cite{Helgaker2000}. The full configuration interaction quantum Monte Carlo (FCIQMC) method \cite{Booth2009} is a particular protocol to sample the ground state eigenvector stochastically and sparsely, allowing one to obtain accurate energies and properties of many-body problems with huge Hilbert-space dimension (e.g.~up to $10^{108}$ in Ref.~\cite{Shepherd2012}). In such problems neither the matrix nor the full ground state vector could be stored in computer memory at one time. FCIQMC is classified as a projector quantum Monte Carlo approach \cite{Umrigar2015} aimed at approximating the ground state, although variations of the FCIQMC protocol have been introduced to compute excited states \cite{Blunt2015b}, finite-temperature problems \cite{Blunt2015a} with density matrices \cite{Blunt2014},
transcorrelated non-hermitian Hamiltonians with three-body interactions \cite{Jeszenszki2020,Cohen2019a,Dobrautz2018},
real-time evolution \cite{McClean2015,Guther2017a}, and driven-dissipative problems for open quantum systems \cite{Nagy2018}. The method itself is fairly young and under active development \cite{Booth2014,Petruzielo2012,Blunt2015,Cleland2010,Ghanem2019,Blunt2018,Holmes2016,Neufeld2019,Greene2019}.

During an FCIQMC simulation, the quantum state is represented, at any one time, by a set of discrete walkers or \textit{psi}-particles, which have to be represented in computer memory \cite{Booth2009}. Storing these walkers presents the primary memory requirement for large-scale simulations. While the storage of walkers that share the same configuration can be optimized, and walker storage can be distributed over many compute nodes in a high-performance computing environment \cite{Booth2014}, the total number of walkers that can be used is limited by the memory hardware resources. On the other hand it is usually desirable or even required to work with as large walker numbers as possible. Large walker numbers may be required to mitigate the sign problem by enabling annihilation of walkers with different signs  \cite{Booth2009,Spencer2012}, to eliminate a systematic bias if the initiator approximation is used \cite{Cleland2010,Ghanem2019}, to eliminate the population control bias \cite{Vigor2015},  or simply to reduce statistical noise in estimators for desired observables like the ground state energy.

\begin{figure}
 \includegraphics[width=0.5\textwidth]{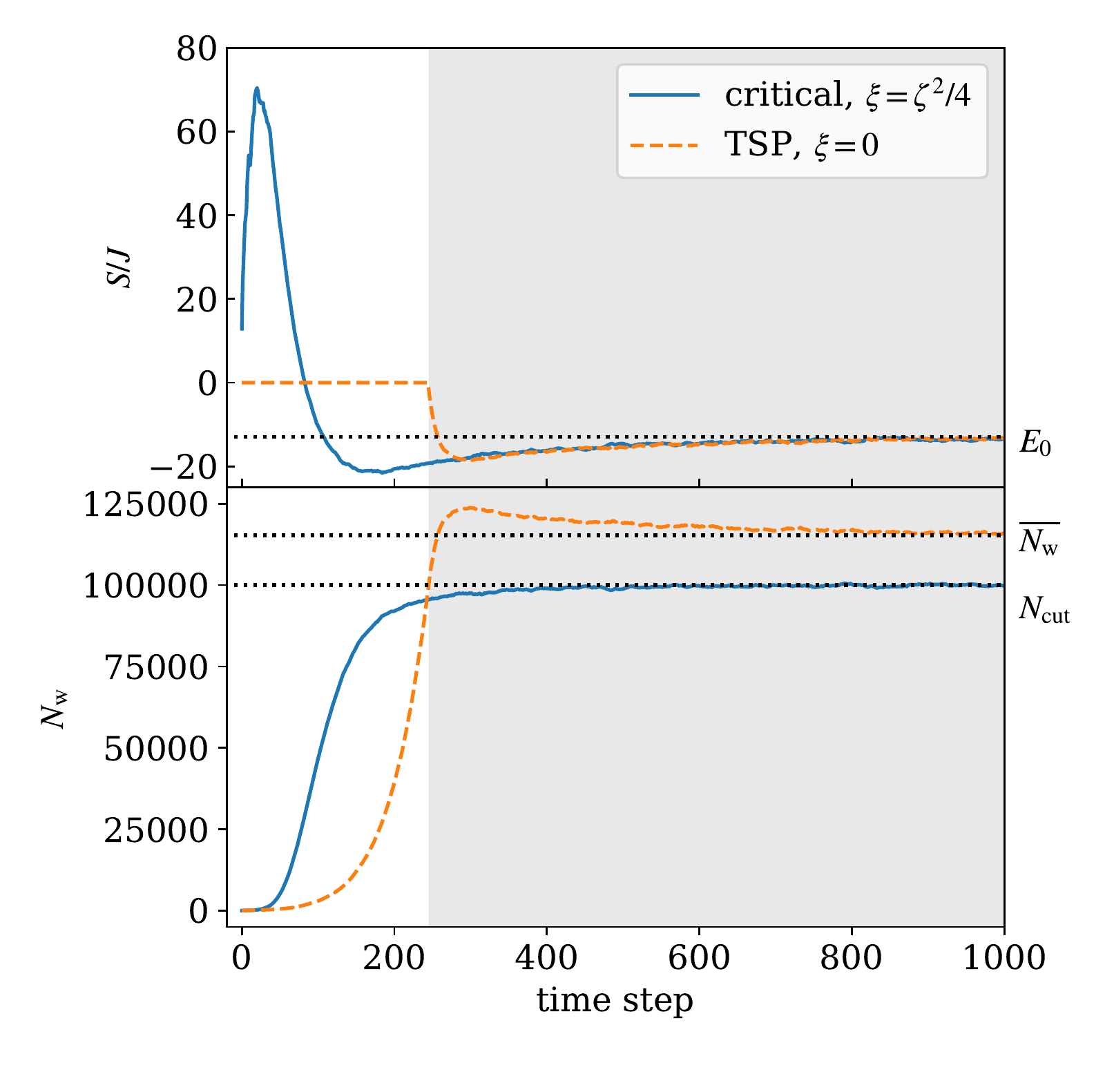}
 \caption{Walker population control in the two-stage procedure (TSP) used in the original FCIQMC \cite{Booth2009}, where $\xi=0$, and the single-stage procedure of Eq.~\eqref{eq:fciqmc-shift-new} with the restoring force set to critical damping ($\xi = \zeta^2/4 $).
The top panel shows the shift and the dotted line indicates the value of the exact ground state energy $E_0$. The walker number is shown in the bottom panel.
Without the restoring force, the maximum of the walker number reaches $N_\mathrm{w, max} \approx 124,000$ and later equilibrates to $ \overline{N_\mathrm{w}} \approx 115,000$ indicated by the upper dotted line. The value of $N_\mathrm{cut} = N_\mathrm{t} = 10^5$ is
indicated by the lower dotted line.
We have chosen a time step of $\delta\tau=0.001 J^{-1}$ and set the damping parameter to $\zeta = 0.08$ for both procedures.
Parameters of the Bose--Hubbard model are $U/J=6$, $N=M=20$.}
 \label{fig:overshoot}
\end{figure}

In the original FCIQMC algorithm \cite{Booth2009} the walker number is controlled by an energy shift parameter $S$ and  there are two stages of walker population dynamics during the time evolution through iterations: In the first stage the shift is kept at a constant value $S_0$ and the number of walkers $N_\mathrm{w}$ is allowed to grow exponentially up to a threshold  value $N_\mathrm{cut}$. In the second stage the shift is updated dynamically to counteract the growth of the walker number, controlled by a damping parameter $\zeta$.  In the long-time limit the walker number will settle to fluctuate around a mean value
\begin{align} \label{eq:Nwbar}
\overline{N_\mathrm{w}} \approx N_\mathrm{cut} \exp\left[\frac{(S_0-E_0)\delta\tau}{\zeta}\right] ,
\end{align}
as is shown in this work, where $\delta\tau$ is a time-step parameter.
The final mean value $\overline{N_\mathrm{w}}$ is larger than the preset value for $N_\mathrm{cut}$ and depends on the \emph{a priori} unknown value of the ground state energy $E_0$.
Moreover, it is possible to get overshoots, where the walker number significantly exceeds both the target value  $N_\mathrm{cut}$ and the final mean value $\overline{N_\mathrm{w}}$ at intermediate times before settling to fluctuate around $\overline{N_\mathrm{w}}$.  An example of such behavior is shown in Fig.~\ref{fig:overshoot}, where the dashed orange  lines show the evolution of the shift $S$ (top) and walker number $N_\mathrm{w}$ (bottom) in the two-stage procedure of Ref.~\cite{Booth2009}.
The fact that the maximum and final average number of walkers are not directly determined by the parameters of the simulations complicates the planning of computational resources and may be met with an over-allocation of resources or requires elaborate estimation or multi-step procedures. A tighter control of the walker number with a pre-defined target value is clearly desirable.


In this work we propose a modified population control procedure for FCIQMC by introducing an additional ``forcing'' term characterized by a 
new
parameter $\xi$ and a target walker number $N_\mathrm{t}$. The new term represents a restoring force that  will push the walker population towards the target value $N_\mathrm{t}$.
The behavior of the new procedure is shown in Fig.~\ref{fig:overshoot} by the full blue lines, where the shift is adjusted from the beginning of the simulation and the walker number quickly equilibrates around the pre-defined value $N_\mathrm{t}$.
Analyzing the population dynamics by a simplified scalar model reveals that the logarithm of the walker number follows the dynamics of a damped harmonic oscillator equilibrating at $N_\mathrm{w} = N_\mathrm{t}$. The walker number control in
the original FCIQMC  \cite{Booth2009} corresponds to the special case without restoring force
or $\xi=0$. We argue that the optimal choice for the new parameter $\xi$ is the value $\xi = \zeta^2/4$ for critical damping in the  scalar model, which removes it as a free parameter from the algorithm.   
A more detailed discussion of Fig.~\ref{fig:overshoot} will follow in Sec.~\ref{sec:overshoots}.

An important aspect of the original protocol is the possibility to diagnose the sign problem by closely examining the walker number growth \cite{Booth2009,Vigor2015}. The sign problem is sometimes referred to as the fermion sign problem, as it is inevitable for some fermionic Hamiltonians \cite{Troyer2005} but it also appears as a dynamical sign problem in time-evolution problems \cite{Cohen2015}, or even for bosonic Hamiltonians (see Appendix \ref{sec:BHMmom}). In FCIQMC it manifests itself by the lack of sign coherence in the walker population when the walker number is insufficient. The sign problem is overcome by the annihilation of positive and negative walkers once the walker number has surpassed a threshold \cite{Booth2009,Vigor2015}. Observing a plateau of stagnant growth during the walker growth stage with constant shift has been used to semi-automatically detect this important threshold value \cite{Shepherd2014}. In this work we introduce a quantity called growth witness, constructed from the logarithmic growth rate of the walker number and the instantaneous shift. The growth witness is able to detect the annihilation threshold by a tell-tale maximum feature. Detection of the annihilation threshold is possible while the population growth is dynamically controlled even though a plateau in the walker number may not be present, thus obviating the need for an uncontrolled growth phase.

%
%

This paper is organized as follows: After introducing the FCIQMC algorithm with modified walker population control in Sec.~\ref{sec:fciqmc} we briefly introduce the model and computational details in Sec.~\ref{sec:simdetails}.
In Sec.~\ref{sec:scalar} we derive  a scalar population dynamics \color{black}
model for the walker dynamics and discuss the solutions with and without forcing term
 in a simplified differential equation formulation as well as stability thresholds for the discrete-time dynamics. \color{black}
Walker number overshoots and long-time limits are discussed in  Sec.~\ref{sec:overshoots} where we also derive Eq.~\eqref{eq:Nwbar}.  
The growth witness is introduced in Sec.~\ref{sec:plateau} where we discuss the annihilation plateau and the detection of the annihilation threshold. \color{black}
Section \ref{sec:fluctuations} deals with fluctuations in the equilibrium phase of the simulation and examines the influence of the damping parameter and parameters of the model Hamiltonian before concluding in Sec.~\ref{sec:conclusion}. Details of the Bose--Hubbard model are presented in Appendices \ref{sec:BHM}  (real space) and \ref{sec:BHMmom} (momentum space).
The effect of delaying the shift update is analyzed and summarized in Appendix \ref{sec:delayed_A}. We finally show that introducing the additional ``forcing'' term to the population control mechanism has no effect on the intrinsic population control bias of FCIQMC in Appendix \ref{sec:pcb}.
\color{black}
%


\section{Walker population control in FCIQMC}\label{sec:fciqmc}

In the configuration interaction approach, the many-body quantum state or wave function is represented by a vector $\mathbf{c}$ composed of coefficients that give the signed weights of individual Fock states, or configurations. Correspondingly, the quantum Hamiltonian is represented by a matrix $\mathbf{H}$. The FCIQMC algorithm is based on the iterative equation describing the update of the coefficient vector $\mathbf{c}^{(n)}$ at the $n-$th time step in discrete time steps $\delta \tau$:
 \begin{equation}
 \mathbf{c}^{(n+1)} = [\mathds{1}+\delta\tau(S^{(n)}\mathds{1}-\mathbf{H})]\mathbf{c}^{(n)} .\label{eq:fciqmc-C}
\end{equation}
Here $\mathds{1}$ represents the unit matrix,
$S^{(n)}$ is the energy shift at time step $n$, and we use units where $\hbar=1$.  The iteration prescription of Eq.~\eqref{eq:fciqmc-C}, if performed exactly, will make the vector $\mathbf{c}^{(\infty)}$ proportional to the ground state eigenvector, the dominant eigenvector of $-\mathbf{H}$. The procedure can be understood either as a repeated matrix-vector multiplication and variant of the power method, or as executing Euler steps of the discretized imaginary time evolution in the Schrödinger  equation. The actual FCIQMC algorithm is a stochastic procedure involving discrete walkers that is aimed to solve Eq.~\eqref{eq:fciqmc-C} on average \cite{Booth2009,Booth2014}. The coefficient vector $\mathbf{c}^n$ at any one time is made up of integer numbers and its one-norm $\lVert \mathbf{c} \rVert_1 \equiv \sum_i \vert c_i \vert$ is interpreted as the number of walkers $N_\mathrm{w}$. Representing the coefficients as integers and controlling the total number  $N_\mathrm{w}= \lVert \mathbf{c} \rVert_1$ allows for a sparse representation of the coefficient vector, where only non-zero elements have to be stored in memory. This is particularly efficient in the typical scenario where the dimension of the linear space is much larger than the number of walkers. The walker number $N_\mathrm{w}$ thus controls the demand for physical memory consumption of an FCIQMC simulation. While the notion of integer walkers was relaxed to include fractional walkers, and floating-point coefficients in a limited subspace of Hilbert space in the context of semi-stochastic FCIQMC \cite{Petruzielo2012,Blunt2015}, the basic principle remains the same. The walker number (defined by the one-norm $N_\mathrm{w}= \lVert \mathbf{c} \rVert_1$) still controls the memory consumption in addition to demands for representing the deterministic space .

In order to control the number of walkers, the original FCIQMC algorithm  \cite{Booth2009} proposed a two-stage procedure
\begin{subequations}\label{eq:fciqmc-shift-orig}
\begin{align}
 S^{(n)}  & = S_0 &\textrm{Stage 1}, \label{eq:fciqmc-shift-orig-a}\\
S^{(n+A)} &= S^{(n)} - \frac{\zeta}{A \delta\tau}\ln\left(\frac{N_\mathrm{w}^{(n+A)}}{N_\mathrm{w}^{(n)}}\right) &\textrm{Stage 2} .
\label{eq:fciqmc-shift-orig-b}
\end{align}
\end{subequations}
During Stage 1 the shift is kept at a constant value $S_0$, usually set to the lowest diagonal matrix element of $\mathbf{H}$, in order to let the walker number grow from a small starting value until it reaches a threshold value $N_\mathrm{cut}$. After the threshold is reached, Stage 2 is activated and the shift is updated every $A$ time steps according to Eq.~\eqref{eq:fciqmc-shift-orig-b},
where $\zeta$ is a dimensionless damping parameter and parameter ranges of $\zeta = 0.05$--0.1 and $A=5$--$10$ are proposed in Ref.~\cite{Booth2009}.
The shift update procedure counteracts any exponential growth of the walker number caused by Eq.~\eqref{eq:fciqmc-C} by lowering the shift, and conversely also counteracts exponential damping by raising the shift $S$. An equilibrium is reached when the coefficient vector $\mathbf{c}^{(n)}$ is proportional to the ground state vector $\mathbf{c}_0$ and the shift equal to the ground state energy, $S=E_0$. A steady state is reached where neither the walker number nor the shift changes. In the typical case where the stochastic realization of  Eq.~\eqref{eq:fciqmc-C} introduces noise, both the shift and the walker number will fluctuate around their equilibrium values. The equilibrium value of the walker number is not predefined in the procedure of Eq.~\eqref{eq:fciqmc-shift-orig} but depends on the initial conditions.

Motivated by the walker control mechanism in diffusion Monte Carlo, where an energy control parameter is adjusted when the walker number deviates from a target value  \cite{Umrigar1993}, we propose the following modified shift-update procedure
\begin{align}
S^{(n+A)} = S^{(n)} - \frac{\zeta}{A \delta\tau}\ln\left(\frac{N_\mathrm{w}^{(n+A)}}{N_\mathrm{w}^{(n)}}\right) -
\frac{\xi}{A \delta\tau}\ln\left(\frac{N_\mathrm{w}^{(n+A)}}{N_\mathrm{t}}\right),\label{eq:fciqmc-shift-new}
\end{align}
which reduces to the original update equation \eqref{eq:fciqmc-shift-orig-b} for $\xi=0$. The dimensionless parameter $\xi$ represents a forcing strength and $N_\mathrm{t}$ is the target walker number. It is easily seen that under steady-state conditions the last two terms must vanish and thus the walker number will equilibrate at the target walker number $N_\mathrm{t}$.
In contrast to the original FCIQMC procedure where the shift is updated only after a threshold number of walkers has been reached, the new shift update procedure of Eq.~\eqref{eq:fciqmc-shift-new} can be used from the start of a simulation, even if the initial walker number is very different from the desired final number $N_\mathrm{t}$.

In the remainder of this paper
we set $A=1$  for simplicity, unless specified 
otherwise. I.e.~the shift is updated in every time step.
Our experience with using larger values of $A$ indicates that there is little benefit in such a choice. The arguments and numerical result are summarized in Appendix \ref{sec:delayed_A}.
\color{black}

\section{Simulation details}\label{sec:simdetails}

All FCIQMC simulations for this paper were done on the Bose--Hubbard model  \cite{Fisher1989}, which is relevant to ultra-cold atom experiments in optical lattices \cite{Gross2017,Jaksch1998,Greiner2002,Bakr2010}.
We use a one-dimensional configuration with periodic boundary conditions. The Hamiltonian and details of the model are summarized in Appendix \ref{sec:BHM}. The simulations were performed with the library \texttt{Rimu.jl} written in the programming language Julia by the authors for FCIQMC with bosonic many-body models
  \cite{rimu2020}.
  Other
implementations of FCIQMC targeting quantum chemical applications as well as the (Fermi) Hubbard model and spin models are publicly available \cite{Guther2020,Spencer2019}.

While we use the original integer walker-number FCIQMC algorithm of Ref.~\cite{Booth2009}
in all numerical examples in this work, the proposed walker population (norm) control mechanism is equally applicable to other variants and flavors of FCIQMC including the initiator approach \cite{Cleland2010}, semistochastic FCIQMC \cite{Petruzielo2012}, and fast randomized iteration schemes \cite{Greene2019}.  All simulations shown in the main part of the paper were conducted with large enough walker numbers to suppress the population control bias known to exist in diffusion Monte Carlo-like schemes \cite{Umrigar1993,Vigor2015} to levels smaller than our stochastic error bars. 
Quantifying the population control bias with reduced walker number we found no detectable influence of 
the modified shift-update procedure of Eq.~\eqref{eq:fciqmc-shift-new} at the level of our stochastic errors, 
as shown in Appendix \ref{sec:pcb}.

\section{Scalar model of walker population dynamics}\label{sec:scalar}

We will further analyze the effects of the shift-update procedure on the dynamics of the walker number with a simple scalar model. In order to motivate the model, let us assume that the coefficient-vector update of Eq.~\eqref{eq:fciqmc-C} is performed exactly and that the coefficient vector is already proportional to the ground state  with $\mathbf{c}^{(n)} = N_\mathrm{w}^{(n)} \mathbf{c}_0$. Here $ \mathbf{c}_0$ is an eigenvector of the matrix-vector equation $ \mathbf{H} \mathbf{c}_0 = E_0 \mathbf{c}_0$  with energy eigenvalue $E_0$ and $N_\mathrm{w}^{(n)}$ the walker number in the $n$th time step.  Then Eq.~\eqref{eq:fciqmc-C} reduces to a scalar update equation for the walker number
\begin{align} \label{eq:wn}
N_\mathrm{w}^{(n+1)} = [1+\delta\tau(S^{(n)}-E_0)]N_\mathrm{w}^{(n)} .
\end{align}
Together with the shift update equation \eqref{eq:fciqmc-shift-new}, it defines the walker number dynamics in discrete time.

\subsection{Population dynamics in continuous time}\label{sec:continuoustime}

Aiming at approximating this dynamics by a differential equation, we introduce a time variable $t = n \delta\tau$ and a new variable $x$ for the logarithm of the ratio between the momentary and the target walker number
\begin{align}\label{eq:def-x}
x^{(n)} = \ln \frac{N_\mathrm{w}^{(n)}}{N_\mathrm{t}} \longrightarrow x(t) = \ln \frac{N_\mathrm{w}(t)}{N_\mathrm{t}} .
\end{align}
The shift update equation \eqref{eq:fciqmc-shift-new} can be written in terms of $x$ as
\begin{align}  \label{eq:logShiftUpdate}
\frac{S^{(n+1)} - S^{(n)}}{\delta\tau} = - \frac{\zeta}{\delta\tau}\frac{x^{(n+1)} - x^{(n)}}{\delta\tau} - \frac{\xi}{\delta\tau^2}x^{(n)} ,
\end{align}
which is a finite difference approximation of the differential equation
\begin{align} \label{eq:dSdt}
\frac{dS}{dt} = - \frac{\zeta}{\delta\tau}\frac{dx}{dt} - \frac{\xi}{\delta\tau^2}x(t) .
\end{align}

After rearranging the walker number equation \eqref{eq:wn}, it is seen to represent a finite difference approximation to the logarithmic time derivative of the walker number
\begin{align} 
 S^{(n)}-E_0  =\frac{N_\mathrm{w}^{(n+1)} - N_\mathrm{w}^{(n)}}{\delta\tau N_\mathrm{w}^{(n)}}  \approx  \frac{d \ln N_\mathrm{w}}{dt} = \frac{dx}{dt}  ,
\end{align}
which yields the differential equation
\begin{align} \label{eq:dxdt}
 \frac{dx}{dt} = S(t) - E_0 .
\end{align}
Equations  \eqref{eq:dSdt} and \eqref{eq:dxdt} form a set of coupled first order ordinary differential equations, which determine the time evolution of $x(t)$ and $S(t)$. By eliminating $S(t)$, the equations can further be combined into a single second order differential equation
for $x$
\begin{align} \label{eq:DHO}
 \frac{d^2 x(t)}{dt^2} + \frac{\zeta}{\delta\tau}\frac{dx(t)}{dt} + \frac{\xi}{\delta\tau^2} x(t) = 0 .
\end{align}
This is the well-known differential equation for the damped harmonic oscillator.
Here, ${\zeta}/{\delta\tau}$ represents a damping coefficient and ${\xi}/{\delta\tau^2}$ the force constant of a restoring force.

\subsection{Walker number dynamics with forcing}

The general solution of the differential equation for the damped harmonic oscillator \eqref{eq:DHO} can be written as
\begin{align}
x(t) = a  e^{-\frac{t}{T_+}} + b e^{-\frac{t}{T_-}} ,
\end{align}
with arbitrary constants $a$ and $b$ whose values are determined by the initial conditions.
The two independent solutions $x_\pm(t) = \exp(-t/T_\pm)$ will both decay to zero in the long-time limit.
The solutions for the time constant are
\begin{align}\label{eq:HOtimescales}
T_\pm =  \frac{\delta\tau}{2\xi}\left(\zeta\pm\sqrt{\zeta^2 - 4\xi}\right).
\end{align}
Depending on the value of the discriminant $\zeta^2 - 4\xi$ we can distinguish three cases corresponding to overdamped, critical and underdamped behavior.
If $\zeta^2 > 4\xi$, the time constants are both real-valued and both fundamental solutions show exponential damping. This is the overdamped case. In the underdamped case of $\zeta^2 < 4\xi$, the square root has imaginary solutions and both fundamental solutions are products of an oscillating component and an exponential damping factor.

The case of critical damping with
\begin{align} \label{eq:criticalxi}
4\xi={\zeta^2} ,
\end{align}
is of particular interest since it is the point at which the damping is the fastest. The critical damping time is given by
\begin{align} \label{eq:Tc}
T_\mathrm{c} = \frac{\delta\tau}{\sqrt{\xi}} = \frac{2}{\zeta} \delta\tau.
\end{align}
Since the exponential ansatz only provides a single fundamental solution of the second order differential equation, another independent solution has to be found. It can be easily checked that a second independent solution is $t \exp(-t/T_\mathrm{c})$. The general solution in the critical damping case is then given by
\begin{align}
x(t) = (a + b  t) e^{-\frac{t}{T_\mathrm{c}}} .
\end{align}
Note that the parameters $\zeta$ and $\xi$ are dimensionless and determine the decay time scale in units of $\delta\tau$. I.e.\ the number of time steps until the solution decays,
$T_\mathrm{c}/\delta\tau$, is dimensionless and
independent of the size of the time step $\delta\tau$.

\begin{figure}
 \includegraphics[width=0.5\textwidth]{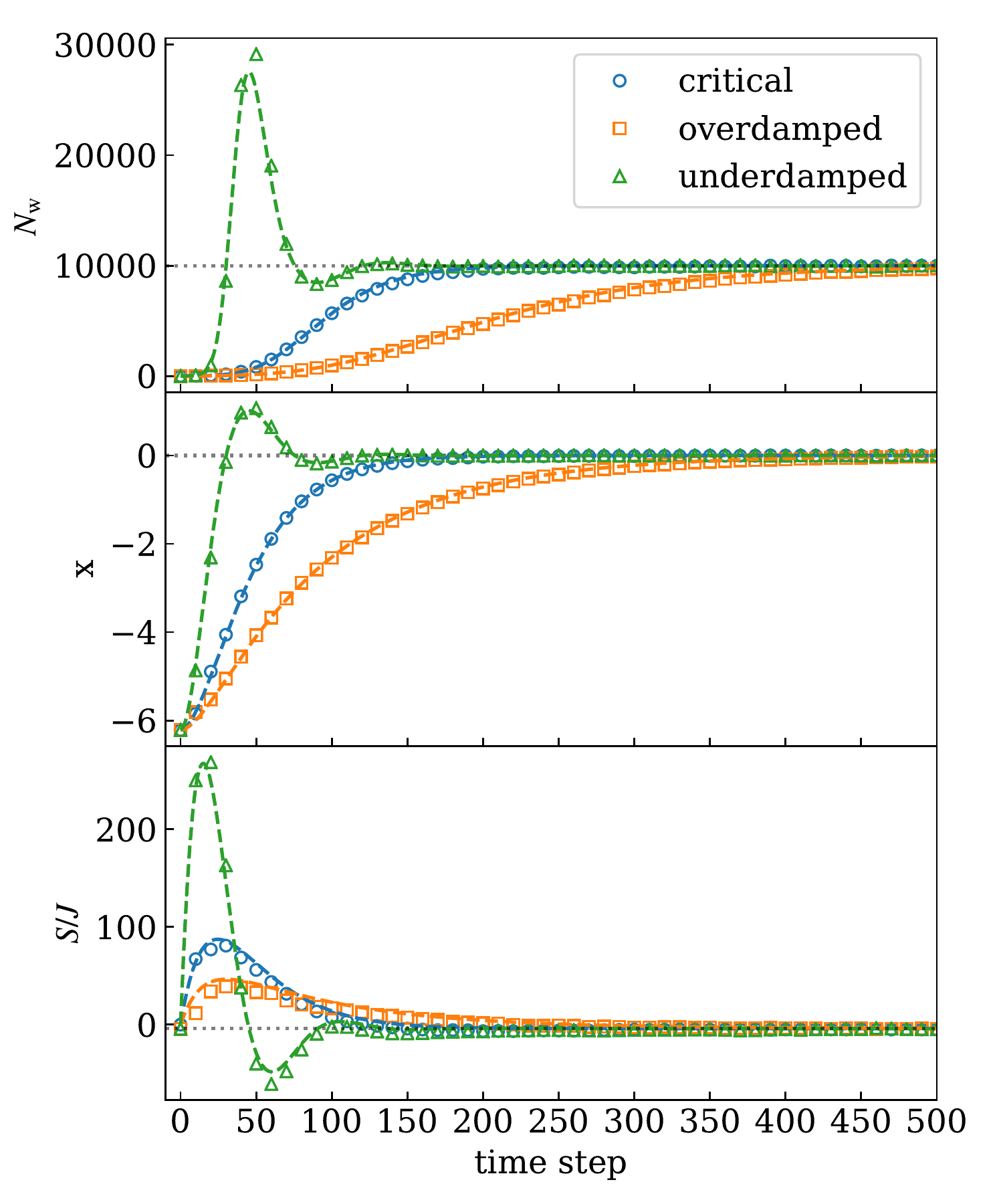}
 \caption{Walker population dynamics in FCIQMC with the shift-update procedure of Eq.~\eqref{eq:fciqmc-shift-new} (symbols) for the Bose--Hubbard model compared with the scalar model of Eq.~\eqref{eq:DHO} (dashed lines). The top panel shows the evolution of the walker number $N_\mathrm{w}$ for three different damping scenarios and the middle and bottom panels show the logarithm $x=\ln (N_\mathrm{w}/N_\mathrm{t})$, and the shift $S$, respectively.
 We used $\zeta = 0.08$ and set  $\xi = 0.0016, 0.0008, 0.0064$ representing the critical, overdamped, and underdamped regimes, respectively. The initial conditions were $N_\mathrm{w}=20$ and $S=0$ at $t=0$, and the target walker number was set to $N_\mathrm{t} = 10,000$.
The parameters of the Bose--Hubbard model are  $M=N=6$ and $U/J=6$. Other parameters used are $\delta\tau=0.001J^{-1}$, and $A=1$.
The dashed lines show the asymptotic values $N_\mathrm{t}=10^4$, ${x} = 0$, and $E_0=-4.0J$ in panels 1, 2, and 3, respectively.
The simulation data is only shown at every tenth time step for clarity.}
 \label{fig:damping}
\end{figure}

Figure \ref{fig:damping} shows how the analytic solutions of the scalar model \eqref{eq:DHO} match FCIQMC simulations of the Bose--Hubbard model very well, demonstrating that underdamped, critical, and overdamped walker number dynamics can be achieved with the new update procedure of Eq.~\eqref{eq:fciqmc-shift-new}.
The time evolution of the walker number is given by $N_\mathrm{w}(t) = N_\mathrm{t} e^{x(t)}$ according to Eq.~\eqref{eq:def-x}.
In the steady-state (long-time) limit, the solution becomes time-independent with $x=0$, or  $N_\mathrm{w} \to N_\mathrm{t}$. The time evolution of the shift is obtained from Eq.~\eqref{eq:dxdt}. In the long-time limit the left hand side vanishes and $S \to E_0$.

\subsection{Walker number dynamics without forcing}

In the original FCIQMC of Ref.~\cite{Booth2009},
after reaching Stage 2 of the  two-stage procedure \eqref{eq:fciqmc-shift-orig},
the evolution of shift and walker number experience damped motion without restoring force ($\xi=0$) and thus no predefined equilibrium exists. In this case it is more convenient to write the differential equation in terms of
\begin{align}
s(t) \equiv S(t) - E_0 = \frac{dx}{dt} ,
\end{align}
which describes the deviation of the shift from the equilibrium value. Combining Eqs.~\eqref{eq:dSdt} and \eqref{eq:dxdt} we then obtain
\begin{align}
\frac{ds(t)}{dt} + \frac{\zeta}{\delta\tau} s(t) = 0 .
\end{align}
This is a simple damping equation with solution
\begin{align}\label{eq:s(t)}
s(t) = [S(0)-E_0] e^{-\frac{t}{T_\mathrm{d}}} ,
\end{align}
with the damping time
\begin{align}\label{eq:Td}
T_\mathrm{d} = \frac{\delta\tau}{\zeta} .
\end{align}
The time dependence for the walker number follows from Eq.~\eqref{eq:def-x}:
\begin{align}
N_\mathrm{w}(t) = N_\mathrm{t} e^{x(t)}
\end{align}
Choosing the time axis to start at $t=0$ when entering Stage 2 of Eq.~\eqref{eq:fciqmc-shift-orig} where $N_\mathrm{w}(0)=N_\mathrm{cut}$ and $S(0)=S_0$, the exponential expression for the walker number $N_\mathrm{w}(t)$ can be expressed  as
\begin{align} \label{eq:damp_orig}
N_\mathrm{w}(t) = N_\mathrm{cut}  \exp\left\{(S_0-E_0) T_\mathrm{d}\left[1- \exp\left(-\frac{t}{T_\mathrm{d}}\right)\right] \right\} .
\end{align}
Taking the long-time limit $t\to\infty$ and substituting Eq.~\eqref{eq:Td} for $T_\mathrm{d}$ we obtain Eq.~\eqref{eq:Nwbar} for the final walker number.

Note that the time evolution described by Eq.~\eqref{eq:damp_orig} is monotonously growing or decaying depending on the sign of $S_0-E_0$. In the standard procedure the initial shift is larger than the final ground state energy, in order to induce walker growth during Stage 1, and thus the scalar model predicts further growth during Stage 2  to the final larger value of Eq.~\eqref{eq:Nwbar}.
The damping time $T_\mathrm{d}$ of Eq.~\eqref{eq:Td} is smaller by a factor of 2 compared to the critical damping time ${T_\mathrm{c}}$ of the damped harmonic oscillator, Eq.~\eqref{eq:Tc}, at the same value of $\zeta$. But this faster damping comes with the cost of reaching a final walker number that depends on the \emph{a priori} unknown value of the ground state energy $E_0$.

\subsection{Population dynamics in discrete time}\label{sec:discretedynamics}

The damped harmonic oscillator differential equation \eqref{eq:DHO} obtained in the continuous-time limit is intuitive and provides much insight. However, it does not capture all aspects of the discrete-time population dynamics described by Eqs.~\eqref{eq:fciqmc-shift-new} and \eqref{eq:wn}.
The discrete-time dynamics will follow closely the differential equation  when the relevant time scales of the damped harmonic oscillator of Eq.~\eqref{eq:HOtimescales} are large compared to the discrete time step $\delta \tau$, i.e.~$\xi, \zeta\ll1$, $\xi/\zeta \ll 1$. Outside of this regime we expect the discrete time dynamics to deviate from the differential equation model.

We can study the discrete time dynamics by treating
the system of equations \eqref{eq:fciqmc-shift-new} and \eqref{eq:wn} as a two-dimensional iterated nonlinear map in the dynamical variables $N_\mathrm{w}^{(n)}$ and $S^{(n)}$.  Assuming $\zeta>0$ and $\xi>0$, it is easily verified that the single fixed point of this iterated map is $N_\mathrm{w}^\mathrm{fp} = N_\mathrm{t}$ and $S^\mathrm{fp} = E_0$. This is completely consistent with the fixed point $x(\infty)=0$ of the damped harmonic oscillator equation \eqref{eq:DHO}.
While the fixed point of the differential equation model is always stable, a standard linear stability analysis \cite{Tel2006} reveals that the fixed point of the iterated map is a stable attractor only if 
\begin{align}\label{eq:stability}
2\zeta + \xi < 4 ,
\end{align}
and is unstable otherwise.
If we set the parameter $\xi$ to the critical damping value of $\zeta^2/4$ as per Eq.\ \eqref{eq:criticalxi}, the condition for stability becomes
\begin{align} \label{eq:stablezeta}
\zeta < 4(\sqrt{2} -1) \approx 1.66 .
\end{align}
%

In the region of stability the fixed point is a spiral attractor if $\zeta < 2\sqrt{\xi} -\xi$ and a node attractor if  $\zeta > 2\sqrt{\xi} -\xi$. For small $\xi$ this is asymptotically equivalent to the condition $ \xi > \zeta^2/4$ for underdamped motion of the harmonic oscillator but for  larger $\xi$ and $\zeta$ the boundary between underdamping (spiral attractor) and overdamping (node attractor) shifts to larger values of $\xi$. The time scale for approaching the fixed point becomes smallest at $\zeta=\xi=1$ for the linearized map where it reaches a single time step $\delta \tau$.
Since the basin of attraction shrinks for the larger $\xi$ values, we nevertheless propose to fix the restoring force coefficient $\xi$ to the critical value $\zeta^2/4$ of Eq.\ \eqref{eq:criticalxi} from the differential equation model.

For the value $\xi=0$ as used in the two-stage procedure a separate stability analysis for the shift-update equation yields the stability condition $\zeta < 2$.

Numerical results concerning the stability of the full FCIQMC iterations and considerations about the optimal choice of $\zeta$ will be discussed in Sec.~\ref{sec:dampingparameter}.

\section{Walker number dynamics in FCIQMC}

\subsection{Final walker number in two-stage procedure}\label{sec:overshoots}

In a real FCIQMC simulation the walker number will fluctuate due to updating the walker number with the complicated and noisy
evaluation of Eq.~\eqref{eq:fciqmc-C}. Even without forcing ($\xi=0$) these fluctuations do not lead to a drift in walker number but instead the walker number is seen to fluctuate around a stable average. This can be understood from the logarithmic update equation \eqref{eq:fciqmc-shift-orig-b}, which is evaluated exactly during the simulation. The update equation \eqref{eq:fciqmc-shift-orig-b} can be re-written in terms of the initial conditions  as
\begin{align}
S^{(n+A)} &= S_0 - \frac{\zeta}{A \delta\tau}\ln\left(\frac{N_\mathrm{w}^{(n+A)}}{N_\mathrm{cut}}\right) ,
\end{align}
which reveals that the value of the shift at any time during the FCIQMC simulation depends only on the initial conditions and the instantaneous walker number but not on the details of fluctuations at intermediate times. Taking the average over many time steps
we obtain
\begin{align}\label{eq:lnbar}
\overline{S} = S_0 - \frac{\zeta}{\delta\tau}\overline{\ln\left(\frac{N_\mathrm{w}}{N_\mathrm{cut}}\right) } .
\end{align}
Replacing the average shift $\overline{S}$ in the long-time limit with the exact ground state energy $E_0$ and approximating the average of the logarithm by the logarithm of the average (with an error $\mathcal{O}[\mathrm{Var}({N_\mathrm{w}}/{N_\mathrm{cut}})]$), we, once again, obtain Eq.~\eqref{eq:Nwbar}.
Note that Eq.~\eqref{eq:lnbar} is an exact result that does not rely on the assumptions of the scalar model and fully includes the effects of a noisy simulation. This means, in particular, that the expression \eqref{eq:Nwbar} for the final walker number is valid for FCIQMC with forceless damping even in situations where the scalar model is not sufficient to fully capture the dynamical evolution of the walker number.



In some cases we see initial growth and overshooting of the walker number before decaying to the long term limiting value  as e.g.~in Fig~\ref{fig:overshoot}. Such non-monotonous behavior of $N_\mathrm{w}(t)$ is not captured by the scalar model solution of Eq.~\eqref{eq:damp_orig}, which predicts monotonous growth.
In Fig~\ref{fig:overshoot} the walker number for the two-stage procedure grows rapidly at the beginning of Stage 2 until saturating at a maximum on a time scale that is consistent with the damping time $T_\mathrm{d}\approx 12.5 \delta \tau$. On the same time scale the shift decays to a minimum value, where it matches the value of the shift obtained with the restoring-force (single-stage) procedure. A further equilibration of the shift to the final value $E_0$ then happens at a much longer time scale over hundreds of time steps for both procedures.
We attribute this behavior to the necessary equilibration of the walker distribution to better represent the ground state vector $\mathbf{c}_0$. This mechanism is not captured by the simplified Eq.~\eqref{eq:wn}, which formed the starting point of the scalar model analysis.
During this period of slow equilibration, the walker number follows the slowly changing shift adiabatically according to Eq.~\eqref{eq:lnbar} for the two-stage procedure without forcing.


The evolution of the walker number for the critical-damping update procedure seen in Fig~\ref{fig:overshoot} is very different though, as here the $\xi$ term forces the walker number back to the target walker number $N_\mathrm{t}$ on the time scale $T_\mathrm{c} = 25\delta\tau$. This time scale is again much faster than the time scale of rearranging the walker population, which affects the slowly changing average of the shift. The equilibration process of the walker population  for FCIQMC in large Hilbert spaces was recently discussed in Ref.~\cite{Neufeld2019a}.
%
%

\begin{figure}
 \includegraphics[width=0.5\textwidth]{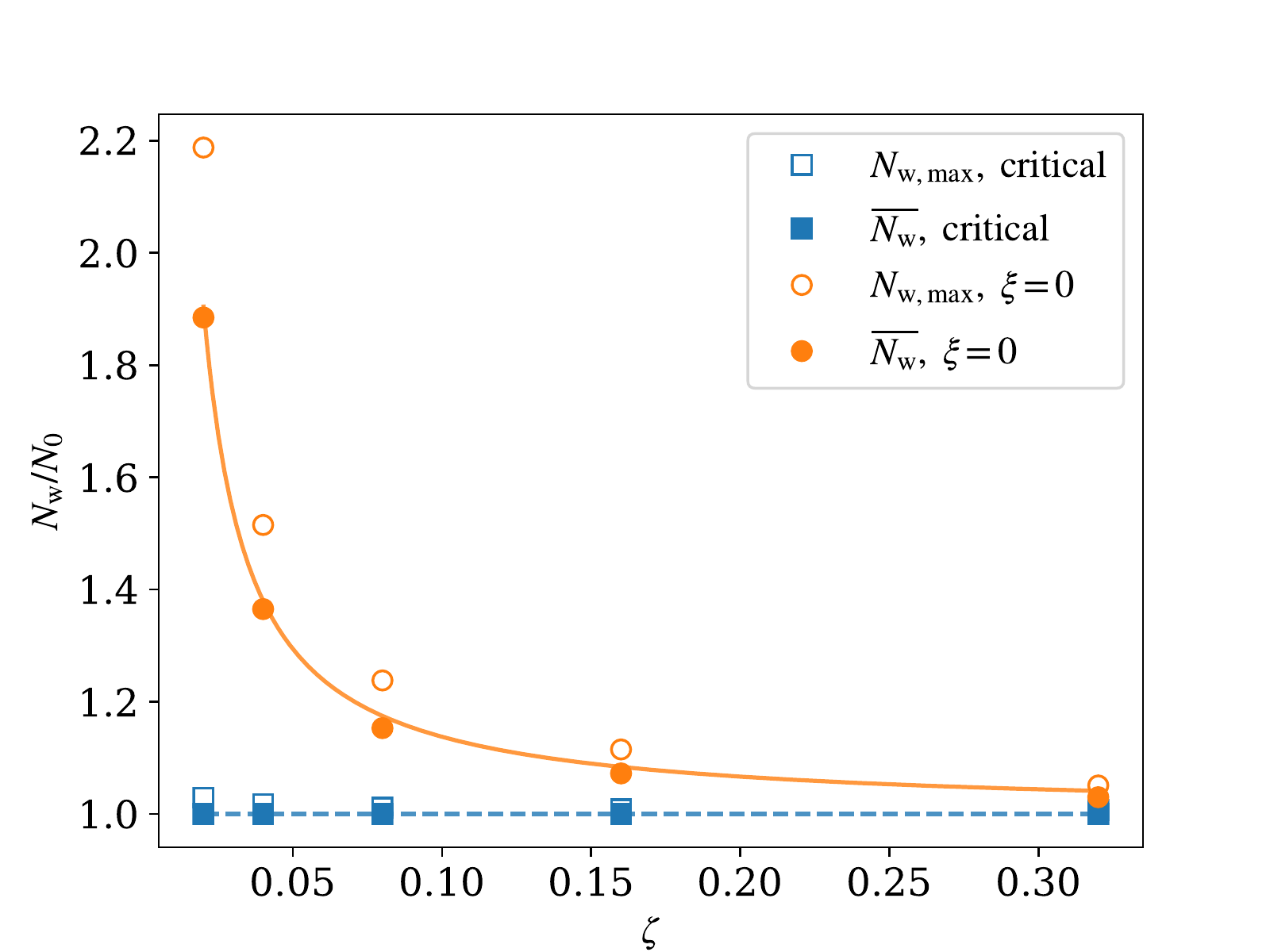}
 \caption{Final walker number and overshoot as function of the damping parameter $\zeta$.
 The maximum walker number reached during a simulation with  ${N_\mathrm{w,max}}$ (empty markers), and the long-time average $\overline{N_\mathrm{w}}$ (filled markers) with $10^6$ time steps are shown for both the single-stage critical damping ($\xi = \zeta^2/4$) and the two-stage procedure  without restoring force ($\xi=0$) and $N_\mathrm{cut} = N_0$ with $N_0 = 10^5$.
 The yellow solid line shows the prediction for $\overline{N_\mathrm{w}}$ without restoring force from Eq.~\eqref{eq:Nwbar}. The blue dashed line shows the prediction $\overline{N_\mathrm{w}}\approx N_\mathrm{t}$ for the single-step procedure with critical forcing where $\xi = \zeta^2/4$ and $N_\mathrm{t} = N_0$. The corresponding data
for the single-stage procedure
is close to the prediction and reveals the superior walker number control and avoidance of overshoots compared to the unforced walker number control.
The parameters of the Bose--Hubbard model are  $M=N=20$ and $U/J=6$, and $\delta\tau=0.001 J^{-1}$. $E_0\approx -12.88 J$
was obtained from long time average of the shift.
} \label{fig:zeta-os}
\end{figure}

Next, we examine the dependence of the overshoot and final walker number on the damping parameter $\zeta$ in Fig.~\ref{fig:zeta-os}. Shown are both  the final average walker number  $\overline{N_\mathrm{w}}$ as well as the maximum number reached during the simulation ${N_\mathrm{w,max}}$, which indicates the overshoot and is the relevant number for computer memory resources.
We find that the exponential dependence predicted by Eq.~\eqref{eq:Nwbar}  captures the results from FCIQMC simulations of the Bose--Hubbard model very well.
Equation \eqref{eq:Nwbar} for $\overline{N_\mathrm{w}}$  (supported by Fig.~\ref{fig:zeta-os}) then suggests that the increase in walker number beyond the threshold value $N_\mathrm{cut}$ can be mitigated by increasing the damping parameter $\zeta$ or decreasing the time step $\delta \tau$. Another possible mitigation strategy would be to set a sequence of intermediate threshold values to smaller values than the final desired value $N_\mathrm{cut}$ and alternate constant-shift and equilibration stages multiple times in order to decrease the energy difference $S_0-E_0$ in  Eq.~\eqref{eq:Nwbar}.
The data from FCIQMC simulations with the one-stage procedure at critical damping shown in Fig.~\ref{fig:zeta-os} demonstrate, however, that both the final and maximum walker number can be very well controlled regardless of the other simulation parameters. Those parameters can then be chosen according to other criteria (\textit{e.g.} larger time steps for faster convergence and better numerical efficiency).

\subsection{Sign problem,  plateau detection, and growth witness}\label{sec:plateau}

\begin{figure*}

\includegraphics[width=1\textwidth]{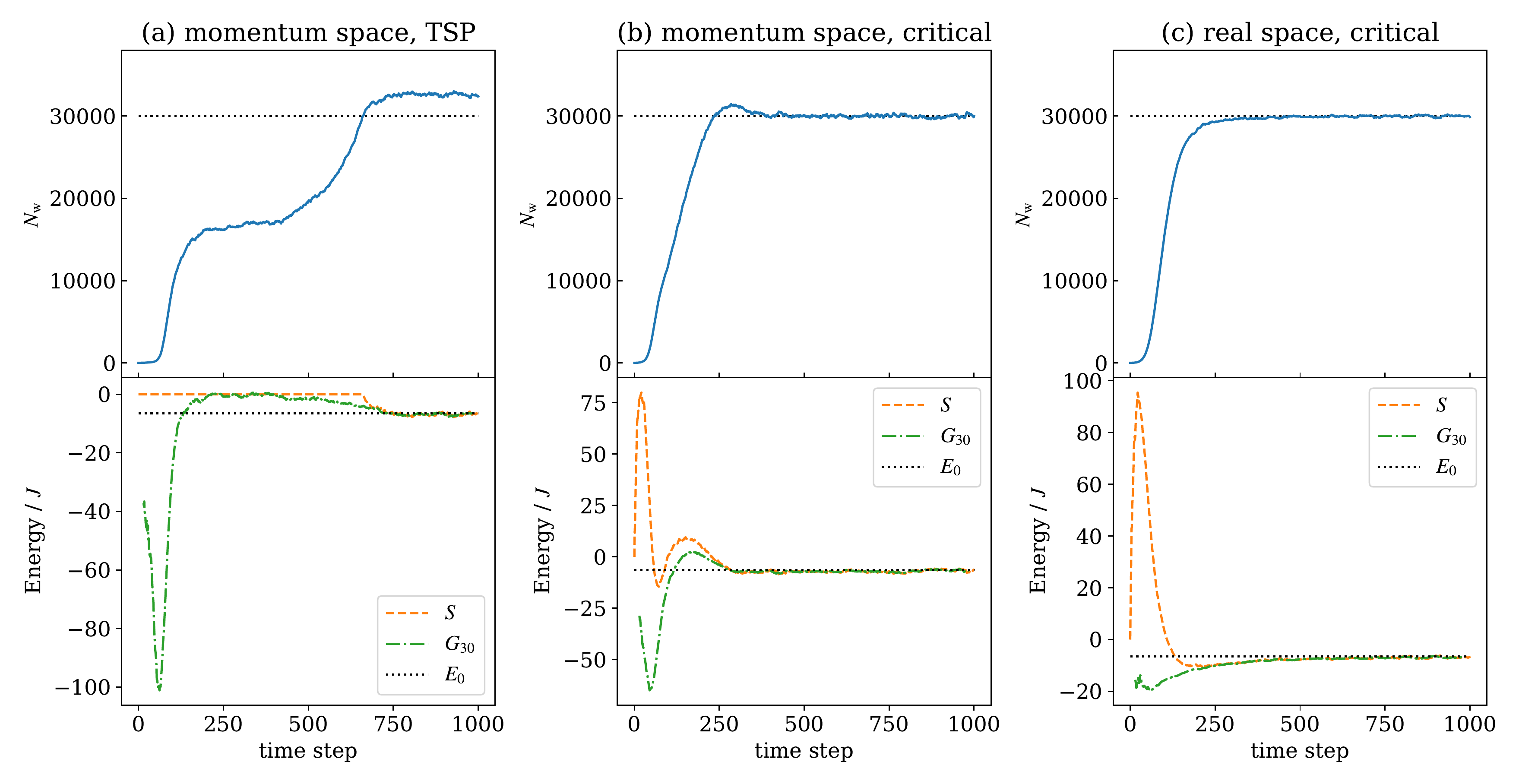}
\caption{\label{fig:plateau}
Dynamics of the walker number $N_\mathrm{w}$ and energy estimators comparing the momentum space Bose--Hubbard model using  (a) the two-stage procedure  (TSP) and (b) walker number control with critical damping ($\xi = \zeta^2/4$) with (c) the real-space Bose--Hubbard model with critical damping.
An annihilation plateau between  growth phases of  $N_\mathrm{w}$ is clearly seen in panel (a), but masked by the population control mechanism in panel (b), and absent in the annihilation-free case of panel (c). The growth witness $G$ provides clear indications of the walker annihilation dynamics with the initial rapid growth phase translating into a low minimum. The onset of the annihilation plateau is represented by crossing the value of $E_0$. $G$ reaches a maximum before settling at the final value $E_0$ in the equilibrated phase of the simulation. These features are  visible in panels (a) and (b), which are based on the same Hamiltonian, whereas the annihilation maximum is absent in panel (c).
The damping parameter is set to $\zeta=0.08$ for all cases.
The parameters of the Bose--Hubbard model are $M=N=10, U/J=6$ and  $\delta\tau=0.001 J^{-1}$. The averaging time scale for $G$ is set to $b = 30$. The exact ground state energy $E_0 = -6.50 J$ was calculated by Lanczos iterations.
}
\end{figure*}

An important situation where the
population dynamics deviates from the simplified model
is when a plateau in the walker number is seen during the constant-shift stage of the two-stage procedure. This is in contrast to the exponential growth that would be expected from the scalar model due to Eq.~\eqref{eq:wn}. The phenomenon was first described in Ref.~\cite{Booth2009} as manifestation of the sign problem in FCIQMC and further analyzed in Ref.~\cite{Spencer2012}. A typical annihilation plateau is seen in Fig.~\ref{fig:plateau} (a) for a calculation using the two-stage procedure on the momentum-space Bose--Hubbard Hamiltonian as described in Appendix \ref{sec:BHMmom}. The initial state was prepared with 20 walkers on the lowest-energy configuration. A phase of rapid growth of the walker number $N_\mathrm{w}$ 
is followed by a stagnant period of almost no growth, which is followed by a second phase of exponential growth. The figure also shows the transition to stage 2 where the walker number is controlled after reaching the predefined value of $N_\mathrm{cut} =$ 30,000. The walker number dynamics for the same Hamiltonian 
and initial state 
with the one-step procedure
is seen in panel (b). It does not show the same plateau due to the forcing term in Eq.~\eqref{eq:fciqmc-shift-new} adjusting the shift as to mandate walker growth before the target $N_\mathrm{t}$ is reached.

In order to disentangle the damped harmonic oscillator dynamics from the annihilation and equilibration dynamics of the walker population it is useful to visit the approximations made in deriving the scalar model. The essential simplification is made when the walker population update of Eq.~\eqref{eq:fciqmc-C} is replaced by the scalar Eq.~\eqref{eq:wn}, which translates into Eq.~\eqref{eq:dxdt} for the amplitude $x = \ln(N_\mathrm{w}/N_\mathrm{t})$. This motivates us to introduce the population growth witness
\begin{align}
G(t) = S(t) - \frac{dx}{dt} ,
\end{align}
which removes the effect of the shift from the  (negative) logarithmic growth rate of the walker number.
It is easily verified from Eq.~\eqref{eq:dxdt} that $G$ manifestly takes the constant value of the ground state energy $E_0$ for the scalar population dynamics model of Sec.~\ref{sec:continuoustime}. Any deviation from this value indicates nontrivial dynamics beyond the scalar model. For actual FCIQMC population dynamics, we define the growth witness at time step $n$ as
\begin{align}
G_b^{(n)} = \bar{S}^{(n)} - \frac{\ln N_\mathrm{w}^{(n+b)} - \ln N_\mathrm{w}^{(n)}}{b \delta\tau} ,
\end{align}
where $b \ge 1$ is the number of time steps for averaging and $\bar{S}^{(n)} = (b+1)^{-1} \sum_{i=n}^{n+b} S^{(n)}$. Averaging of this quantity is useful to smooth out fluctuations because $G$ is related to the derivative of a fluctuating quantity. We found values of $b = 10$--$50$ to be useful.

The growth witness $G_b^{(n)}$ is shown as green dash-dotted line in the lower panels of Fig.~\ref{fig:plateau} along with the exact ground state energy $E_0$ and the shift. It is instructive to interpret the value of the growth witness $G$ during the three stages of walker number growth for the plateau scenario of panel (a). While the shift is held at the constant value zero, $G$ just represents the negative logarithmic derivative of the walker number. During the initial growth phase (before the plateau), the value of $G$ drops to very low values, severely undercutting the actual ground state energy. This can be rationalized by the  population dynamics analysis of Ref.~\cite{Spencer2012}: For low walker numbers, while annihilation of positive and negative walkers is not yet efficient, the FCIQMC iterations of Eq.~\eqref{eq:fciqmc-C} support growing a sign incoherent walker population with a higher growth rate (i.e.~lower $G$) than the actual ground-state eigenvalue $E_0$ would support \footnote{ According to Ref.~\cite{Spencer2012} this is due to the presence of a larger dominant eigenvalue (compared to $-E_0$) of the incoherent transfer matrix. The growth witness measure the negative of this dominant eigenvector during the incoherent phase of walker growth.}. During the plateau phase, annihilation of walkers carrying positive and negative signs becomes efficient  as the walker population becomes large enough to increase the probability for positive and negative walkers to meet on the same configuration.  While the walker number $N_\mathrm{w}$ is stagnant during this phase, the growth witness $G$ rises above the value of $E_0$. Finally, during the second growth phase the overall sign of the walker population is coherent and the population becomes approximately proportional to the actual ground state vector $\mathbf{c}_0$. During this phase, the growth witness $G$ drops from its maximum value to the ground state energy $E_0$ as the assumed relation \eqref{eq:wn} of the scalar model is approximately fulfilled. Note that $G$ is not affected by the onset of walker number control in the second stage of the two-stage procedure because it successfully disentangles the effects of the walker population dynamics from the damping (or forcing) effects of the shift-update procedure.

For the one stage procedure shown in Fig.~\ref{fig:plateau}(b) the growth witness $G$ becomes a  very useful quantity for understanding the population dynamics. By removing the effect of the dynamically adjusted shift from the population dynamics, it shows the same salient features of annihilation-and-growth scenario observed in panel (a):
An initial dip to values much lower than the final asymptote 
is followed by a maximum at values above $E_0$ indicating a phase of efficient annihilation before the value of $G$ finally drops down to the level of $E_0$.
The maximum in $G$ becomes the equivalent of the annihilation plateau and is the tell-tale sign of the emergence of a coherent phase structure in the coefficient vector, which is necessary for overcoming the sign problem. We have found that the walker number at the time when $G$ crosses the asymptotic value $E_0$ from below before the annihilation maximum is a good indicator of the minimum number of walker necessary for the long-term average of the shift to settle at the correct value of the ground state energy $E_0$. It thus replaces the observation of the plateau. If $N_\mathrm{t}$ is set below this value, the shift settles at a value lower than $E_0$ indicating that the sign structure of the (fluctuating) coefficient vector is not fully coherent.
Note that a small temporary overshoot appears in the walker number beyond  $N_\mathrm{t}$, which we interpret as another indicator for reaching a phase of efficient walker annihilation.
The fact that the annihilation maximum appears earlier and for a shorter time than in the two-stage procedure shown in   Fig.~\ref{fig:plateau}(a) can be rationalized by the fact that the walker number grows much earlier  and that the annihilation and equilibration phases will be a function of both total walker number and time. The time scale of walker growth can be adjusted by changing $\zeta$ according to Eq.~\eqref{eq:Tc}. 

Figure \ref{fig:plateau}(c) shows, for comparison the walker number, shift, and the growth witness for the two-step procedure with the real-space version of the Bose--Hubbard Hamiltonian of Appendix \ref{sec:BHM}. Since the real-space Hamiltonian has only non-positive off-diagonal matrix elements, all walkers have the same sign and there are strictly no annihilation events. The real-space Bose--Hubbard Hamiltonian is thus sign-problem-free or stoquastic. The growth witness $G$ is seen (after some initial fluctuations at low walker number) to monotonously increase to the ground state energy from below. The slow approach of  $G$ to the asymptotic value of $E_0$ signifies the convergence time scale of FCIQMC. Importantly, the fact that $G$ never rises above the value of $E_0$ but rather approaches from below indicates the absence of walker annihilation.

\section{Fluctuations in equilibrium}\label{sec:fluctuations}

After an initial phase of dynamics in the walker number and the shift, an equilibrium is reached where the walker number and the shift fluctuate around their long-time average values. The fluctuations originate in the stochastic procedure of evaluating Eq.~\eqref{eq:fciqmc-C}.
During this equilibrium phase the fluctuating coefficient vector $\mathbf{c}^{(n)}$ samples the ground state and the average of the fluctuating shift provides an estimator for the ground state energy.

\begin{figure}
 \includegraphics[width=0.5\textwidth]{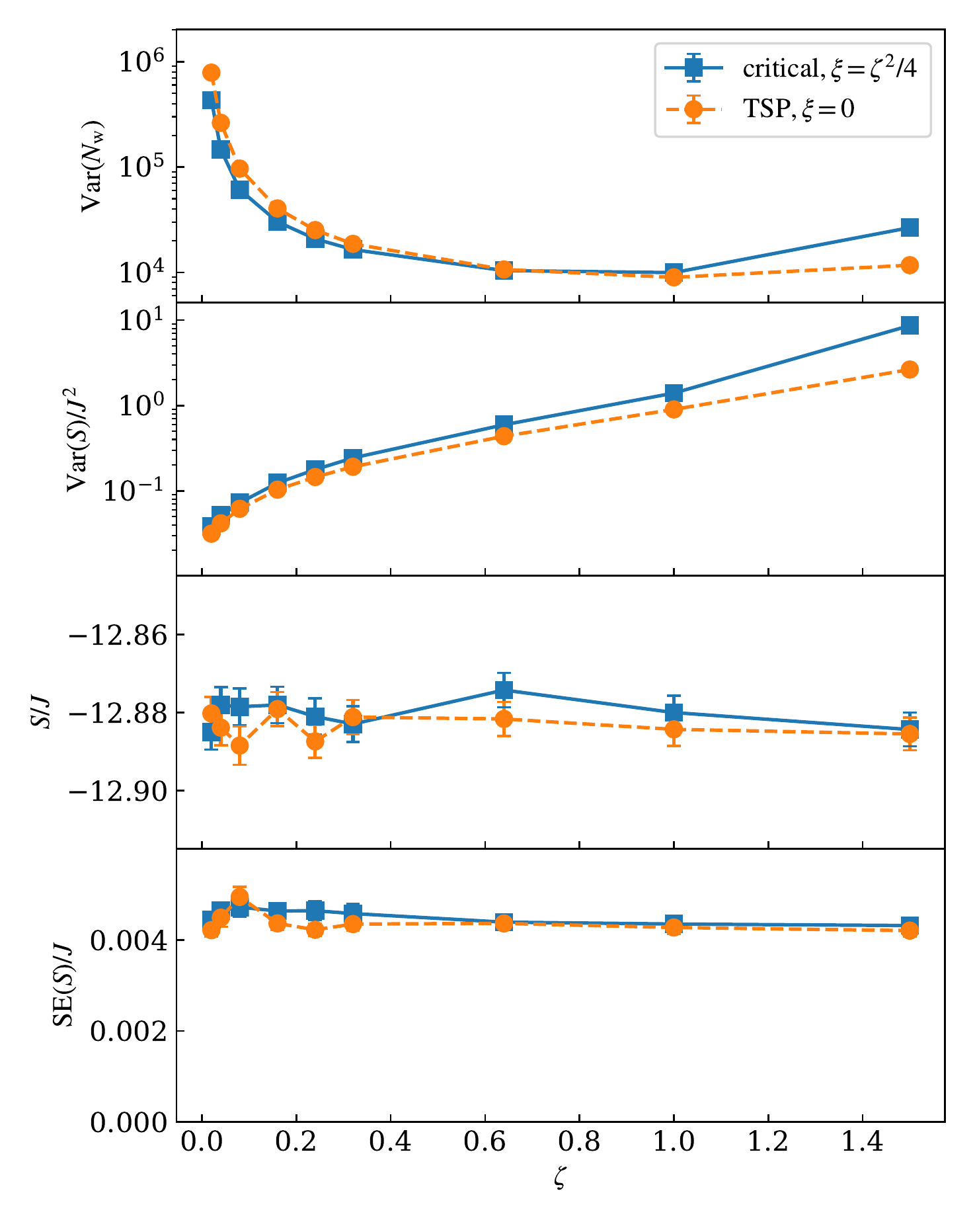}
 \caption{Fluctuating quantities in the equilibrium phase of an FCIQMC simulation as a function of the damping parameter $\zeta$
 with the restoring force set to critical damping ($\xi = \zeta^2/4 $), and for the two-stage procedure (TSP) without restoring force ($\xi=0$).
 The top panel shows the variance of the walker number and the second panel the variance of the shift.
The shift estimator in the third panel was obtained from averaging $10^6$ time steps, and the standard error (shown separately in the bottom panel) was found by blocking analysis.
The same average walker number $\overline{N_\mathrm{w}}=100,000\pm500$ after equilibration was used for both procedures to ensure the results are directly comparable. The parameters of the Bose--Hubbard model are set to $M=N=20, U/J=6$ and  $\delta\tau=0.001 J^{-1}$. Error bars for the variances (calculated from 10 blocks of data with $10^5$ timesteps each) 
and the SE are small and mostly obscured by the markers.
The lines between markers are a guide to the eye. Mind the logarithmic scale for $\mathrm{Var}(N_\mathrm{w})$ and $\mathrm{Var}(S)$. }
 \label{fig:zeta-var}
\end{figure}

\subsection{Effect of the damping parameter}\label{sec:dampingparameter}

Figure \ref{fig:zeta-var} shows how various quantities of interest are affected by the damping parameter $\zeta$ in the equilibrium phase for both the original two-stage procedure without forcing ($\xi=0$) and for the new procedure with the restoring force tuned to the critical value ($\xi = \zeta^2/4$). 
A trade-off can be seen between fluctuations of the walker number, where the variance is suppressed for increasing $\zeta$ (top panel), and the fluctuations in the shift, whose variance grows with increasing $\zeta$ (second panel from top). This is not surprising, since the shift is related to the logarithmic derivative of the walker number per Eq.~\eqref{eq:dxdt} and thus the quantities are conjugate to each other. It is also seen in Fig.~\ref{fig:zeta-var} that the one-stage procedure with restoring force at critical value is more effective in suppressing fluctuations in the walker number
(for $\zeta<0.6$)
than the two-stage procedure without restoring force ($\xi=0$) at the same value of $\zeta$, while the opposite is true for the variance of the shift.
The shift estimator
is shown in the third panel and it can be verified that the obtained values all agree within error bars for all values of the  $\zeta$ and $\xi$ parameters. The error bars
signify the standard error (SE, values shown separately in the bottom panel) obtained from an automated blocking analysis, where the data is de-correlated by blocking transformations \cite{Flyvbjerg1989} and the success of the de-correlation established with the ``M-test'' method by Jonsson \cite{Jonsson2018}. 

\begin{figure}
 \includegraphics[width=0.5\textwidth]{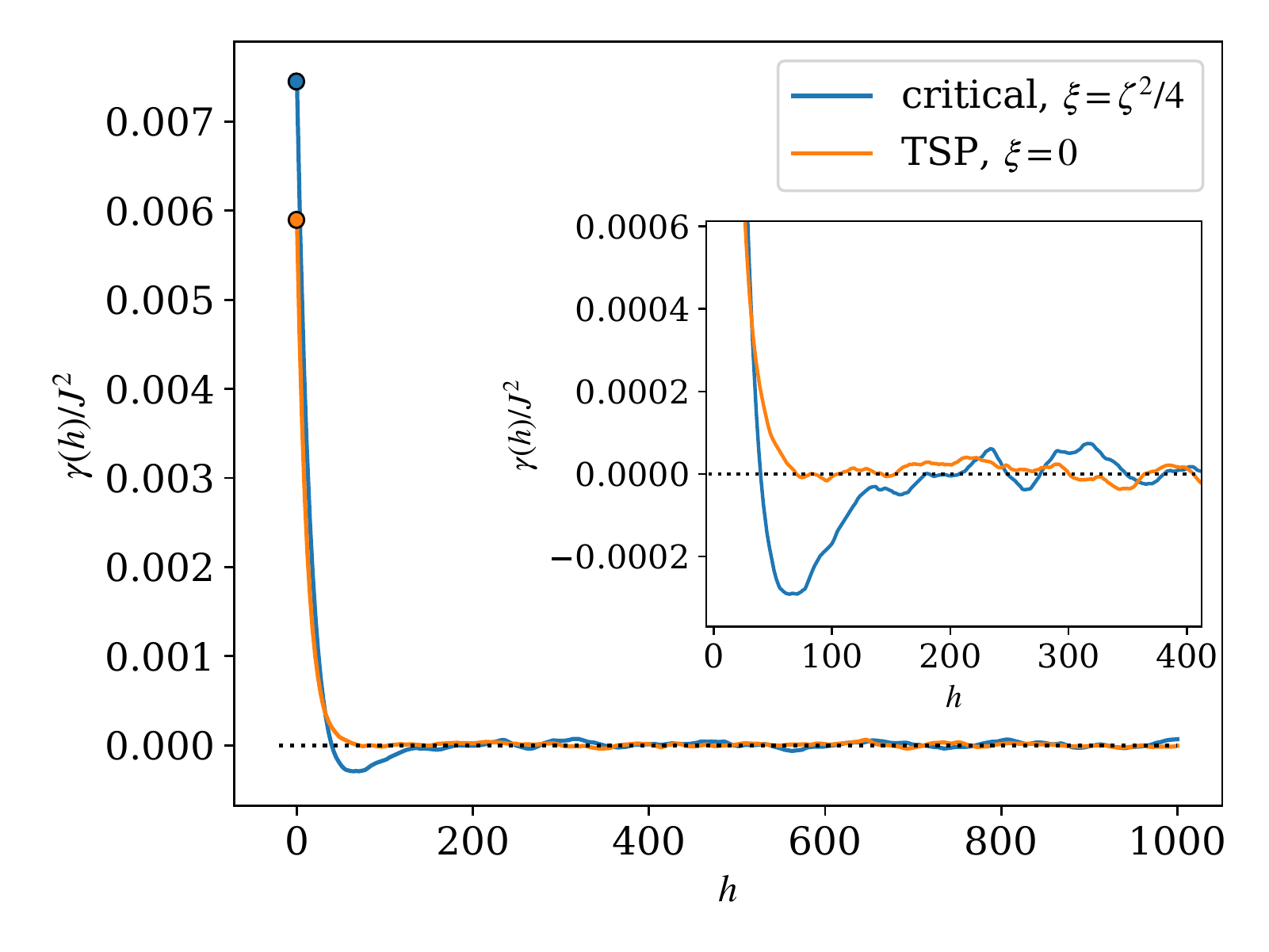}
 \caption{Auto-covariance  of the shift for the
single-stage procedure with the restoring force set to critical damping ($\xi = \zeta^2/4 $) and for the two-stage procedure (TSP) without restoring force ($\xi=0$). The inset shows the details around the zero-crossing point for the critical damping curve. The parameters of the Bose--Hubbard model are set to $M=N=6, U/J=6$.
 The parameters used are $\delta\tau=0.001 J^{-1}$, and $\zeta = 0.08$ for both procedures. The walker population is equilibrated to $\overline{N_\mathrm{w}}=10^5$. Both data sets used to calculate the auto-covariance contain results from $10^6$ time steps. }
 \label{fig:ch}
\end{figure}

The standard error of the shift estimator is an important quantity because it quantifies the quality of the Monte-Carlo simulation
\footnote{
The standard error of the shift over a constant number of time steps reported here is equivalent to the inverse of the statistical efficiency in the language of Ref.~\cite{Greene2019}. Smaller standard error means higher statistical efficiency.
}.
It is  remarkable to see that the same standard error for the shift is obtained for the different values of the damping and forcing parameters (bottom panel, Fig.~\ref{fig:zeta-var}), even though the variances of the shift vary greatly (second panel, Fig.~\ref{fig:zeta-var}).
This fact can be rationalized by considering that the standard error is not only affected by the fluctuations of the shift captured by the variance but also by correlations in the time series. In particular the
squared standard error of the shift estimator (i.e.~the variance of the sample mean of $S^{(n)}$) is obtained from the auto-covariance $\gamma(h)$  by the sum  \cite{Flyvbjerg1989}
\begin{align}
\left[\textrm{SE}(S)\right]^2 = \frac{1}{n_\mathrm{d}}\left[\gamma(0) +2 \sum_{h=1}^{n_\mathrm{d}}\left(1-\frac{h}{n_\mathrm{d}}\right) \gamma(h)\right] ,
\end{align}
where $n_\mathrm{d}$ is the number of data points.
The
auto-covariance of the shift is
\begin{align}
\gamma(h) = \overline{(S^{(n)} - \overline{S})(S^{(n+h)} - \overline{S})} ,
\end{align}
where $\overline{\cdots}$ signifies the sample average over a sufficiently long time series of data. $h$ is a delay in time steps and for $h=0$ the auto-covariance becomes the variance $\gamma(0)= \mathrm{Var}(S)$.
Figure \ref{fig:ch} shows the auto-covariance of the shift for a simulation using the two-stage procedure without restoring force ($\xi=0$) and one with the critical value of the restoring force  ($\xi = \zeta^2/4$) at otherwise identical parameters. While the critically damped simulation has a larger variance of the shift (point for $h=0$ in  Fig.~\ref{fig:ch}) it also features a zero crossing with anticorrelations (negative values) during a significant interval. This makes it possible to yield the same standard error while the variance is different, as seen in Fig.~\ref{fig:zeta-var}.

While a typical range of the damping parameter $\zeta = 0.05$--$0.1$ was proposed Ref.~\cite{Booth2009}, 
the results of Fig.~\ref{fig:zeta-var} suggest that larger values can be used during the equilibrium phase of an FCIQMC calculation without sacrificing the quality (statistical efficiency) of the results. This might be useful if a very tight control of the walker number is necessary.
Consistent with the analysis of Sec.~\ref{sec:discretedynamics}, we find stable damped population dynamics for values of $\zeta \lessapprox 1.6$ (with $\xi = \zeta^2/4$), and unstable dynamics for larger values (including oscillating dynamics for $1.7 \lessapprox \zeta \lessapprox 1.9$). 
If values of $\zeta \gtrapprox 0.5$ together with $\xi$ set to $\zeta^2/4$ or larger are used for growing a walker population from a small size, we find that the target walker number $N_\mathrm{t}$ is reached very quickly (time scale of the order of $\delta\tau$), which precludes the observation of the population growth dynamics and the annihilation maximum in the growth witness $G$. For observing the growth dynamics, we thus used $\zeta < 0.1$. An optimal value may be found when the damping time scale $T_\mathrm{c} = 2\delta\tau/\zeta$ of Eq.~\eqref{eq:Tc} is large compared to $\delta\tau$ but comparable to or smaller than the  FCIQMC convergence time scale on which the walker population becomes a representative sample of the ground state coefficient vector $\mathbf{c}_0$. The latter time scale will depend on the specifics of the Hamiltonian.  


\subsection{Fluctuations in different physical regimes}

\begin{figure}
 \includegraphics[width=0.5\textwidth]{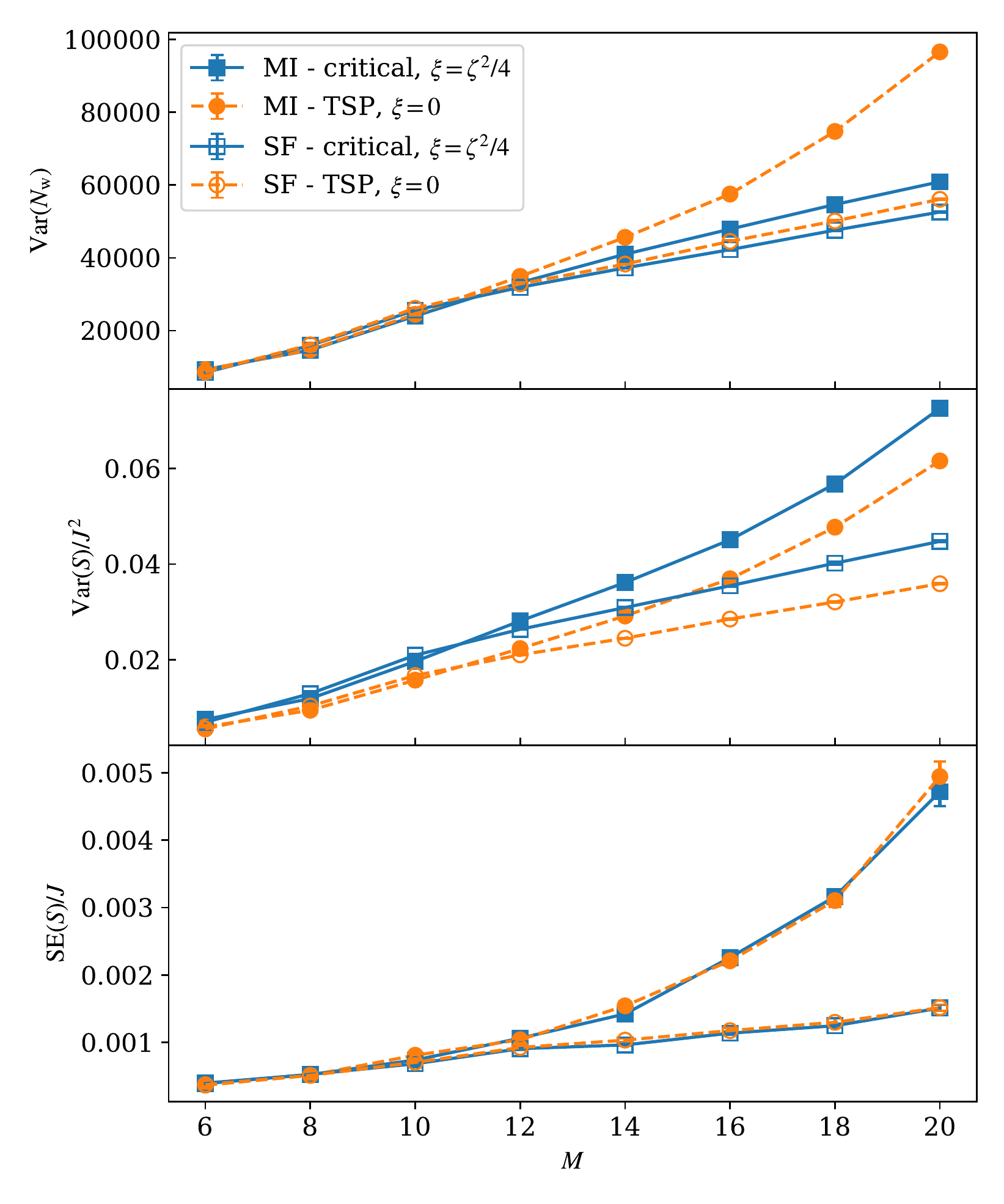}
 \caption{Statistics of the Monte Carlo sampling in the equilibrated regime comparing the population control by the single-stage procedure with the restoring force set to critical damping ($\xi = \zeta^2/4 $) and the two-stage procedure (TSP) without restoring force ($\xi=0$).
 Two distinct physical regimes for the Bose--Hubbard model with unit filling factor are considered: the Mott insulating (MI, $U/J=6$, filled markers) and superfluid (SF, $U/J=1$, empty markers). The parameters used are $\delta\tau=0.001 J^{-1}$, and $\zeta = 0.08$ for both procedures. For all data here the walker population is equilibrated to $\overline{N_\mathrm{w}}=10^5\pm200$.  One million time steps are used to obtain statistically meaningful results. Error bars are mostly within the markers. }
 \label{fig:var-m}
\end{figure}

The Bose--Hubbard model allows us to easily change the parameters of the model to access different physical regimes. The details of the model are discussed in Appendix \ref{sec:BHM}.
Figure \ref{fig:var-m} shows the statistics of the walker number and the shift in the equilibrium phase of an FCIQMC simulation as a function of the system size. The Hilbert space dimension grows rapidly with system size from 462 for $N=6$ particles in $M=6$ lattice sites
to $6.9\times 10^{20}$ for  20 particles in 20 sites
according to Eq.~\eqref{eq:hssize}. With a walker population of $\overline{N_\mathrm{w}}\approx 10^5$, the systems up to $M=10$ have smaller linear spaces than available walkers and thus are well sampled, whereas the Hilbert space dimension rapidly exceeds the walker number for the larger systems.

Figure  \ref{fig:var-m} also shows data for ground states with different interaction strengths: a Mott insulator state with strong interactions $U/J=6$, and a superfluid state with much weaker interactions $U/J=1$. 
The data clearly separate between the Mott insulator and superfluid state for the larger system sizes 
whereas all data are very similar for the smaller systems.


The data for the standard error shows that consistently the two population control procedures give the same quality of Monte Carlo data for the shift estimator, which further confirms the observation made in the previous section.
The rapid increase in the standard error of the shift for the Mott insulator with system size indicates that this state becomes more difficult to sample with the FCIQMC sampling procedure, and this is also reflected by increasing fluctuations of the shift and the walker number. As discussed in more detail in Appendix \ref{sec:BHM}, the Mott insulator state has a single dominant configuration and many small coefficients for other configurations whereas the superfluid state is more evenly spread across Hilbert space, see also Fig.~\ref{fig:MI-SF}. A remarkable difference between the unforced and forced population control procedures is seen in the variance of the walker number in the top panel of  Fig.~\ref{fig:var-m}. A rapid growth with system size for the Mott insulator with the unforced (original FCIQMC) procedure is reduced to a much more moderate increase with the forced procedure. Excessive fluctuations of the shift come with a cost of memory resources that have to be provided for the largest expected demand, and thus it is very desirable to suppress these fluctuations, as the forced shift-update procedure does.

\section{\label{sec:conclusion}Conclusion}

The newly proposed shift-update procedure \eqref{eq:fciqmc-shift-new} with the forcing strength set to the value of critical damping was shown to effectively control the walker number by adjusting it to a pre-defined parameter value $N_\mathrm{t}$. The fluctuations of the walker number are reduced compared to the original procedure without forcing term, while the quality of the Monte Carlo simulation and the shift energy estimator are unaffected by the procedure or the strength of the damping coefficient. Varying the damping coefficient $\zeta$ was shown to have opposite effects on the variances of the shift and particle number.
Values of $\zeta \lessapprox 0.1$, possibly adjusted to the FCIQMC convergence time scale, will be best for observing the walker growth dynamics and detecting an annihilation threshold. However, larger values of $0.5\lessapprox \zeta \lessapprox 1$ can safely be used during the equilibrium phase of an FCIQMC simulation if the strongest suppression of the walker number fluctuations is desired.
The new procedure is simpler than the original one as it removes the necessity for two simulation stages. Moreover it is easy to implement in any FCIQMC code.

An important feature of the walker population dynamics in FCIQMC is the ability to detect the mitigation of the sign problem through efficient walker annihilation. In previous works this was done by detecting an annihilation plateau in the walker growth while the shift is held constant \cite{Booth2009,Shepherd2014}. The detection of the plateau is not straightforward and a histogram analysis of the logarithmic walker number has proven useful in Ref.~\cite{Shepherd2014} but could not be fully automated. 
In this work we have introduced the growth witness $G$,
which displays a maximum at the annihilation plateau. It can be used to detect the threshold walker number at which  annihilation becomes efficient, an overall sign of the coefficient vector emerges, and the sign ``problem'' is successfully mitigated.
While the annihilation plateau in the walker number disappears in the new shift-update procedure (if it is used during the walker growth phase), the growth witness still displays the annihilation maximum and can be used to detect the annihilation threshold. Further research is necessary to show whether the detection of the annihilation threshold via the growth witness $G$ can be successfully automated.


\section*{DATA AVAILABILITY}

The data that support the findings of this study are available from the corresponding author upon reasonable request. The \texttt{Rimu.jl} program  library  is available as an open source project on GitHub \cite{rimu2020}.

\begin{acknowledgments}
We are  grateful to Ali Alavi for enlightening discussions and encouragement.
JB and EP acknowledge the hospitality of the Max-Planck Institute for Solid State Research during the early   development of the FCIQMC code for bosons.
This work
was supported by the Marsden Fund of New Zealand (Contract No.\ MAU1604), from government funding managed by
the Royal Society of New Zealand Te Apārangi. We also acknowledge support by the New Zealand eScience Infrastructure (NeSI) high-performance computing facilities in the form of a merit project allocation and a software consultancy project.
\end{acknowledgments}

\appendix

\section{The Bose--Hubbard model  in real space}\label{sec:BHM}

\begin{figure}
 \includegraphics[width=0.5\textwidth]{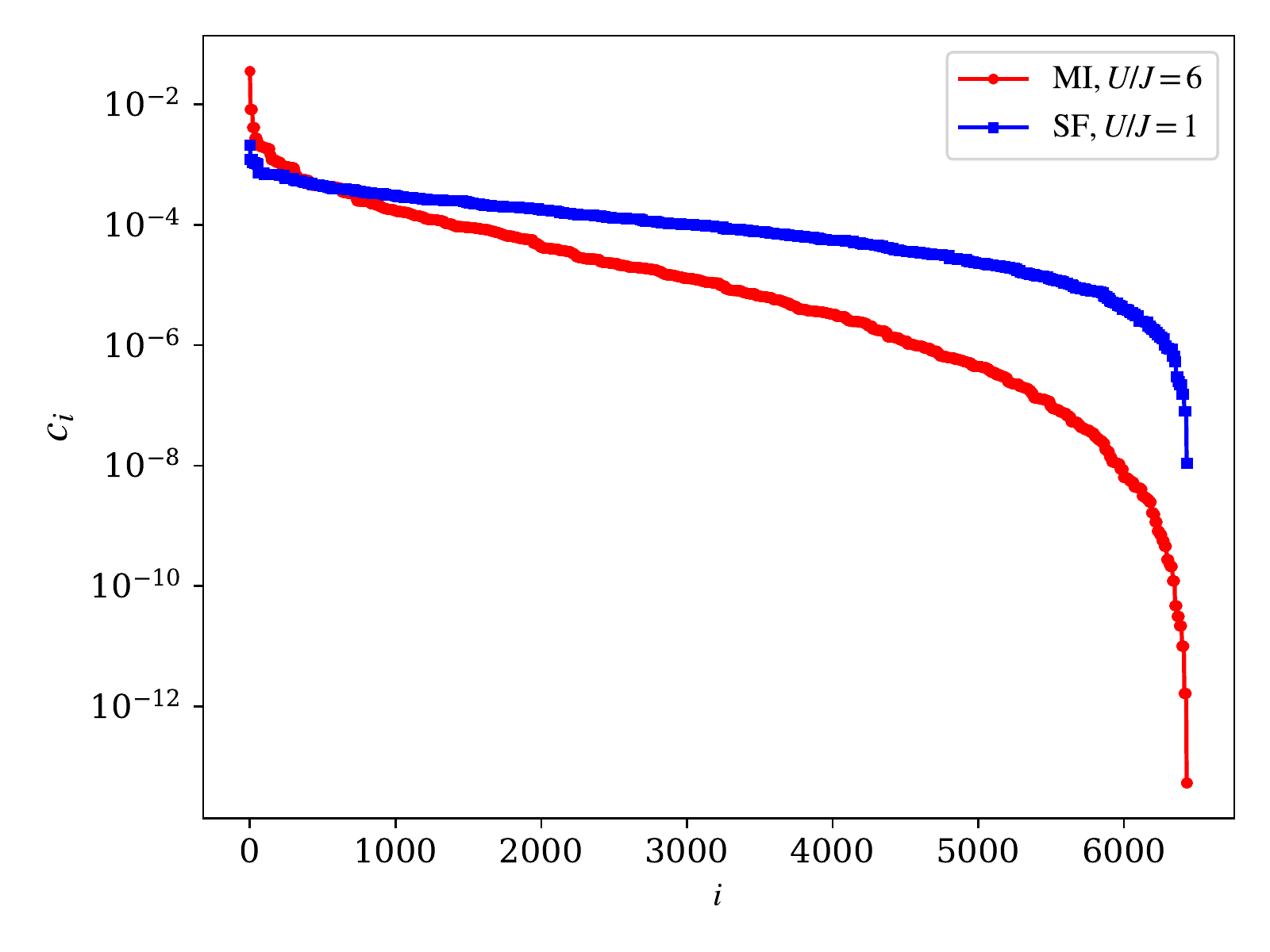}
 \caption{The coefficients of the $\mathbf{c_0}$ vector for the Mott insulating (MI, red dots, $U/J=6$) and superfluid ground state (SF, blue squares, $U/J=1$), ordered by magnitude. The coefficient vector was normalized with the one-norm $\Vert  \mathbf{c_0} \Vert_1 = 1$. The system size is $M=N=8$ and the dimension of Hilbert space is $6435$. }
 \label{fig:MI-SF}
\end{figure}

The Bose--Hubbard Hamiltonian for a one-dimensional chain of $M$ lattice sites is written as
\begin{align}
 H =  -J  \sum_{i=1}^M\left(\hat{a}_i^\dag \hat{a}_{i+1} + \hat{a}_{i+1}^\dag \hat{a}_{i}\right)+ \frac{U}{2} \sum_{i=1}^M \hat{n}_i (\hat{n}_i -1) ,\label{eq:bhm}
\end{align}
where $\hat{a}_i^\dag$ ($\hat{a}_i$) is the creation (annihilation) operator for a particle at site $i$ with bosonic permutation relations $[\hat{a}_i,\hat{a}_j^\dag] = \delta_{ij}$ and $[\hat{a}_i,\hat{a}_j] = 0$, and $\hat{n}_i = \hat{a}_i^\dag\hat{a}_i$ is the number operator. Periodic boundary conditions imply that $\hat{a}_{M+1} \equiv \hat{a}_1$. The total particle number $\hat{N} = \sum_{i=1}^M \hat{n}_i$ is a good quantum number and in our simulation we set it to a fixed value $\hat{N}=N$. The first term in Eq.~\eqref{eq:bhm} represents particle hopping to nearest neighbor sites with hopping strength $J$, and the second term is an on-site interaction with strength parameter $U$.
The Bose--Hubbard model is a non-trivial many-body problem. It has been realized experimentally with ultra-cold atoms in optical lattices \cite{Greiner2002}, with quantum gas microscopes allowing single-atom-level configuration readout \cite{Bakr2010}.

In order to represent the model Hamiltonian as a matrix, we use an occupation number basis (also Fock states, or configurations) in real space
\begin{align}
|n_1, n_2, \ldots, n_M\rangle = \prod_{i=1}^{M}\frac{1}{\sqrt{n_i!}} \left(\hat{a}_i^\dag\right)^{n_i} |\mathrm{vac}\rangle ,
\end{align}
with fixed particle number $N=\sum_{i=1}^M n_i$.
The number of independent basis states with $N$ particles in $M$ lattice sites  and thus the dimension of the matrix $\mathbf{H}$ is
\begin{align}\label{eq:hssize}
\mathrm{dim} = \binom{M+N-1}{N} .
\end{align}
It can be easily adjusted, as $N$ and $M$ are just parameters of the model and the code.

In the thermodynamic limit ($M,N \to \infty$), the one-dimensional Bose--Hubbard model features a quantum phase transition between a Mott-insulating phase characterized by an integer number of particles per lattice site and a gapped excitation spectrum to a gapless superfluid phase \cite{Fisher1989}. While states with non-integer filling factor $N/M$ are always superfluid, the phase transition happens for unit filling $N=M$ at a value of $U/J \approx 3.5$, where larger values correspond to the Mott insulator and smaller values to the superfluid phase.

As all off-diagonal matrix element of the 
real-space
Bose--Hubbard Hamiltonian are non-positive, and thus the matrix is stoquastic \cite{Bravyi2010}, the annihilation of walkers in FCIQMC algorithm will not be triggered. This allows us to bypass the ``annihilation plateau'' and avoid the QMC sign-problem, hence to focus on the dynamics that is solely controlled by the equation of the shift. In Figure \ref{fig:MI-SF} we show the coefficients of the ground state eigenvector $\mathbf{c}_0$ for two specific states of a finite system that are deep inside the Mott-insulating and superfluid regimes, respectively. The Mott-insulating state has a single dominant configuration $\prod_{i=1}^M \hat{a}_i^\dag |\mathrm{vac}\rangle$ in addition to  many small-magnitude coefficients while the superfluid state is much more evenly spread out across Hilbert space.

\section{The Bose--Hubbard model in momentum space}\label{sec:BHMmom}

For the study of the plateau and the sign problem conducted in Sec.~\ref{sec:plateau}, we reformulate the 1D Bose--Hubbard Hamiltonian in momentum space \cite{Zhang2011a},
\begin{align}
 H =  -J  \sum_k\epsilon_k \hat{m}_k + \frac{U}{2M} \sum_{kpqr} \hat{b}_r^\dag \hat{b}_q^\dag \hat{b}_p \hat{b}_k \delta_{r+q,p+k},\label{eq:bhm-mom}
\end{align}
where $\epsilon_k = -2J\cos(k)$ and $\hat{m}_k =   \hat{b}_k^\dag \hat{b}_k$ is the number operator. Single-particle mode operators now refer to plane-wave eigenstates of the lattice momentum
$\hat{b}^\dag_k = M^{-\frac{1}{2}}\sum_{l=1}^M e^{ikl} \hat{a}^\dag_l$, where $k= -\pi + n 2\pi/M$ for even $M$ and $k= -\pi (M+1)/M + n 2\pi/M$ for odd $M$ and $n = 1, \ldots M$. In this formulation, the Hamiltonian is no longer stoquastic when $U>0$ because the off-diagonal matrix elements have a positive sign. Since the Hamiltonian appears with a negative sign in Eq.~\eqref{eq:fciqmc-C}, every spawning event will reverse the sign of a walker. Since spawned walker can arrive at a configuration from different origins with different signs, annihilation events can occur. Evidence for the tell-tale annihilation plateau is seen in Fig.~\ref{fig:plateau}(a).


\section{Effect of the shift update delay $A$}\label{sec:delayed_A}

For all simulations shown in the main part of the paper we set $A=1$ for the delayed shift update in Eqs.~\eqref{eq:fciqmc-shift-orig-b} and \eqref{eq:fciqmc-shift-new}. The parameter $A$ was introduced in Ref.~\cite{Booth2009} where values between 5 and 10 were used. Here, we examine the role of $A$ in the shift update and the population control process in order to determine whether gains in statistical efficiency or savings in computational cost can be made.

In the context of the scalar population dynamics model of Sec.~\ref{sec:scalar}, the effect of $A$ is to increase the size of the effective time step to $A\delta\tau$. Thus the stability boundaries of the discrete-time model [e.g.~Eq.~\eqref{eq:stability}] remain  independent of $A$ (since they do not depend on the time step size) and  the same differential equation \eqref{eq:DHO} is obtained  in the limit $\delta\tau\to 0$.

The effect of the parameter $A$ is to separate the effective time step for the stochastic coefficient vector update in Eq.~\eqref{eq:fciqmc-C} from the shift update of Eqs.~\eqref{eq:fciqmc-shift-orig-b} and \eqref{eq:fciqmc-shift-new}. 
%
%
%
%
With the original two-stage shift update procedure, $A$ does not affect the population dynamics in the walker growth phase as the shift is initially kept constant.
In contrast,  the modified shift update adjusts the shift according to Eq.~\eqref{eq:fciqmc-shift-new} during the walker growth phase in order to achieve damped oscillator motion for the logarithmic walker number. Values of $A>1$ allow walker growth or decay that is exponential in $A$ and generally does not conform with the controlled oscillator motion. This can lead to undesired overshoots or rapid walker number decline. Thus we do not generally recommend using $A>1$ during the walker growth phase.
%
%
%

\begin{figure}

 \includegraphics[width=0.5\textwidth]{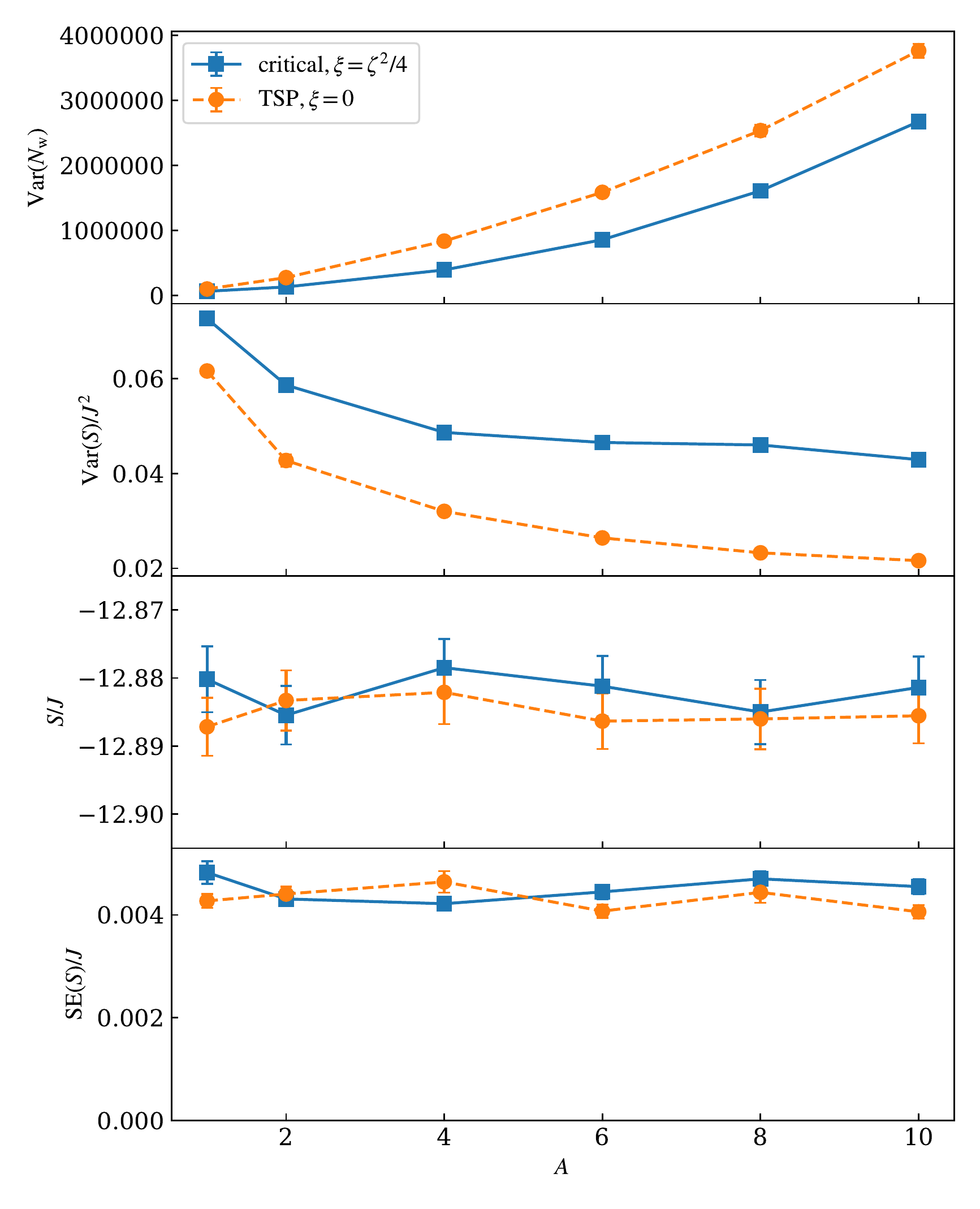}
 \caption{Fluctuating quantities in the equilibrium phase of an FCIQMC simulation as a function of the shift update delay $A$,
 with the restoring force set to critical damping ($\xi = \zeta^2/4 $), and for the two-stage procedure (TSP) without restoring force ($\xi=0$).
 The top panel shows the variance of the walker number and the second panel the variance of the shift.
The shift estimator in the third panel was obtained from averaging $10^6$ time steps, and the standard error (shown separately in the bottom panel) was found by blocking analysis.
The same average walker number $\overline{N_\mathrm{w}}=100,000\pm500$ after equilibration was used for both procedures to ensure the results are directly comparable. The parameters of the Bose--Hubbard model are  $M=N=20$, $U/J=6$ and  $\delta\tau=0.001 J^{-1}$. Error bars for the variances (obtained by evaluating  ten blocks of $10^5$ data points) and the SE each are mostly within the markers.   The lines between markers are a guide to the eye. }
  \label{fig:var-A}
\end{figure}

As the shift serves not only as the population controller but also as an energy estimator, it is relevant to study the role of $A$ in the equilibrium phase. 
Results from an FCIQMC calculation are shown in Fig.~\ref{fig:var-A}.
Both the case of  critical damping and the case of $\xi=0$ show similar trends as a function of $A$. The fluctuations in the shift reduce as $A$ is increased while the variance of the walker number grows significantly. Consistent with what is seen in other parts of this work, the variance of the shift is  smaller with the modified shift-update with critical forcing compared to the original procedure with $\xi=0$, with a trade-off in the increased fluctuations of the shift. Importantly, neither the mean of the shift nor its standard error are significantly affected, which means that the statistical efficiency is constant.

It remains to consider the computational costs. Updating the shift at each time step comes with small constant cost for either procedure of Eq.~\eqref{eq:fciqmc-shift-orig-b} or Eq.~\eqref{eq:fciqmc-shift-new}. Performing the updates at every $A$th step divides this cost by $A$. However, the main cost of the FCIQMC algorithm comes from looping over the coefficient vector and performing spawning operations, which scales with $\mathcal{O}(N_\mathrm{w})$. With our code {\tt Rimu.jl}  \cite{rimu2020}, the CPU runtime deviates by less than 1.3\% for different values of $A$.


\section{Population control bias} \label{sec:pcb}

A (typically) small bias that disappears with increasing walker number is known to affect FCIQMC estimators for observables  \cite{Vigor2015}. Due to the analogy with a conceptually related bias in diffusion quantum Monte Carlo \cite{Umrigar1993} it is known as the ``population control bias'', but it was also termed ``statistical bias'' in Ref.~\cite{Greene2019}. The bias is often difficult to detect and smaller than statistical error bars when large walker numbers are mandated by the annihilation plateau for overcoming the sign problem. In stoquastic problems like the real-space Bose--Hubbard model of Appendix \ref{sec:BHM} no such requirements exist and the bias can be detected by  simply reducing the walker number. Here we examine how the population control bias is affected by the new shift update procedure of Eq.~\eqref{eq:fciqmc-shift-new}.
 
 \begin{figure}
 
\includegraphics[width=0.5\textwidth]{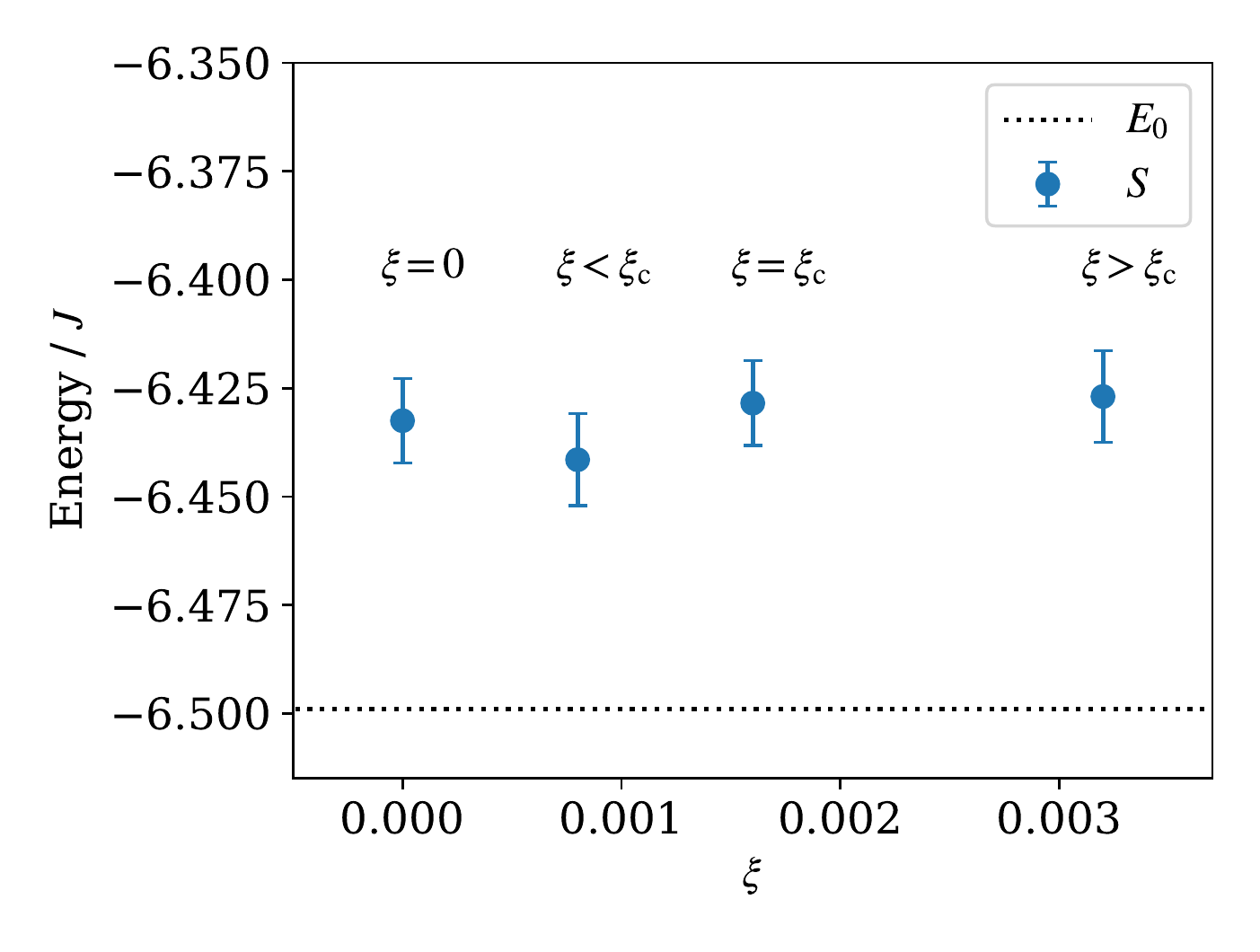}
 \caption{The averaged shift energy estimator displaying the population control bias resolved with a small number of walkers $\overline{N_\text{w}}=1,000$. For comparison, the dimension of Hilbert space is $\mathrm{dim}=$ 92,378. The parameters of the Bose--Hubbard model are  $M=N=10$ and $U/J=6$. The exact ground state energy of the system $E_0 = -6.50 J$ calculated by Lanczos iterations is indicated by the dotted line.
 We used $\zeta = 0.08$ and set $\xi =0$ representing the original FCIQMC shift update procedure, and $\xi = 0.0016, 0.0008, 0.0032$ representing the critical, overdamped, and underdamped regimes, respectively.
 Other parameters used are $\delta\tau=0.001J^{-1}$, and $A=1$. }
 \label{fig:PCB}
\end{figure}

Figure \ref{fig:PCB} shows the population control bias resolved with $N_\mathrm{t} = 1000$ walkers for the shift energy estimator of a real-space Bose Hubbard Hamiltonian with 6 particles in 6 lattice sites. The data points with different values of the forcing constant $\xi$ represent the old shift-update procedure of Eq.~\eqref{eq:fciqmc-shift-orig-b} for $\xi=0$ as well as the underdamped, critically damped, and overdamped regimes, respectively. We find that the population control bias remains unchanged for the different shift-update procedures within our statistical error bars, and conclude that any possible influence of the forcing term on the population control bias is undetectable with our current data.

%
%
%


\bibliography{shift-update.bib,Methods-local,Methods}

\end{document}